\documentclass[sigconf]{acmart}

\usepackage{booktabs} 
\usepackage{subcaption}
\usepackage{longtable}
\usepackage[colorinlistoftodos]{todonotes}
\usepackage{color,soul}
\usepackage{xcolor}

\setcopyright{rightsretained}

\begin{document}
\title{Fault Detection Effectiveness of Source Test Case Generation Strategies for Metamorphic Testing}

\author{Prashanta Saha}
\orcid{}
\affiliation{%
  \institution{School of Computing, Montana State University}
  \city{Bozeman}
  \state{Montana}
  \postcode{59715}
}
\email{p66n633@msu.montana.edu}

\author{Upulee Kanewala}
\authornote{Corresponding author}
\affiliation{%
 \institution{School of Computing, Montana State University}
  \city{Bozeman}
  \state{Montana}
  \postcode{59715}
}
\email{upulee.kanewala@montana.edu}






\renewcommand{\shortauthors}{}

\begin{abstract}
Metamorphic testing is a well known approach to tackle the oracle problem in software testing. This technique requires the use of source test cases that serve as seeds for the generation of follow-up test cases. Systematic design of test cases is crucial for the test quality. Thus, source test case generation strategy can make a big impact on the fault detection effectiveness of metamorphic testing. Most of the previous studies on metamorphic testing have used either random test data or existing test cases as source test cases. There has been limited research done on systematic source test case generation for metamorphic testing. This paper provides a comprehensive evaluation on the impact of source test case generation techniques on the fault finding effectiveness of metamorphic testing. We evaluated the effectiveness of line coverage, branch coverage, weak mutation and random test generation strategies for source test case generation. The experiments are conducted with 77 methods from 4 open source code repositories. Our results show that by systematically creating source test cases, we can significantly increase the fault finding effectiveness of metamorphic testing. Further, in this paper we introduce a simple metamorphic testing tool called "METtester" that we use to conduct metamorphic testing on these methods. 

\end{abstract}

%
%



 \maketitle


\section{Introduction}

 A \emph{test oracle} \cite{Weyukerarticle} is a mechanism to detect the correctness of the outcomes of a program. The \emph{oracle problem} \cite{6963470} can occur when there is no oracle present for the program or it is practically infeasible to develop an oracle to verify the correctness of the computed outputs.
 This test oracle problem is quite frequent especially with scientific software and is one of the most challenging problems in software testing.
Metamorphic testing (MT) technique was proposed to alleviate this oracle problem \cite{Chen:2015:MTS:2819261.2819278}. MT uses properties from the program under test to define  metamorphic relations (MRs). A MR specifies how the outputs should change according to a specific change made into the source input. Thus, from existing test cases (named as source test cases) MRs are used to generate new test cases (named as follow-up test cases). Then the set of source and follow-up test cases are executed on the program under test and the outputs are checked according to the corresponding MRs. The program under test can be considered as faulty if a MR is violated.

Effectiveness of MT in detecting faults  depends on the quality of MRs. Additionally the effectiveness of MT should also rely on the source test cases. Effectiveness of metamorphic testing can be improved by systematically generating the source test cases. Such a systematic approach can reduce the size of the test suite and could be more cost effective. Most of the previous studies in MT have used  randomly generated test cases as source test data for metamorphic testing.
In this study we investigated the effectiveness of line, branch coverage, weak mutation, and random testing for creating source test cases for MT.

Our experimental results show that test cases satisfying weak mutation coverage provide the best fault finding effectiveness. We also have found  that combining one or more systematic source test case generation technique(s) may increase the fault detection ability of MT.




\section{Background}
MT is a property based testing approach which aims to  alleviate the oracle problem. But the effectiveness of MT not only depends on the quality of MRs but also on the source test cases. In this section we briefly discussed MT and source test  generation techniques, line, branch coverage and weak mutation.

\subsection{Metamorphic Testing}
 Oracle problem is one of the biggest challenges in software testing. MT is an effective method to test program that faces oracle problem. MT \cite{Chen:2015:MTS:2819261.2819278}   creates the follow-up test cases from the existing test cases called source test cases. To generate follow-up test cases first we need to identify an appropriate set of MRs that test program under test (PUT) should satisfy. MRs \cite{6319263} are identified based on the properties of the problem domain like the attribute of the algorithm used. We can create source test cases using techniques like random testing, structural testing or search based testing. Follow-up test cases are generated by applying the input transformation specified by the MRs. After executing the source and follow-up test cases on the PUT we can check if there is a change in the output that matches the MR, if not the MR is considered as violated. Violation of MR during testing indicates fault in the PUT. Since MT checks the relationship between inputs and outputs of a test program, we can use this technique when the expected result of a test program is not known.
 
\begin{figure*}[t]
  \includegraphics[scale=0.8]{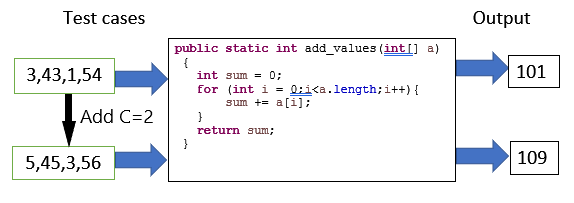}
  \caption{Test Source and follow-up inputs on PUT.}
  \label{fig:10}
\end{figure*}

For example, in figure \ref{fig:10},  a Java method \textit{add\_values} is used to show how  source and follow-up test cases work with a PUT. The \textit{add\_values} method sum up all the array element passed as argument. Source test case, $t = \{3,43,1,54\}$ is randomly generated and tested on \textit{add\_values}. The output for this test case is 101. For this program, when a constant $c$ is added  to the input, the output should increase. This will be used as a  MR to conduct MT on this PUT. A constant value 2 is added to this array to create a follow-up test case $t^{'} = \{5,45,3,56\}$ and then run on the PUT. The output for this follow-up test case is 109. To satisfy this Addition MR the follow-up test output should be greater than the source output. In this MT example, the considered MR is satisfied for this given source and follow-up test cases. 

\subsection{Source Test Case Generation}
To generate source test cases we have used the EvoSuite \cite{Fraser:2011:EAT:2025113.2025179} tool. EvoSuite is a test generation tool that automatically produces test cases targeting a higher code coverage. EvoSuite uses an evolutionary search approach that evolves whole test suites with
respect to an entire coverage criterion at the same time.  In this paper we generated source test cases based on line, branch coverage , weak mutation and random testing. Below we briefly describe the systematic approaches used by EvoSuite to generate them.

\subsubsection{Line Coverage}

In line coverage \cite{Rojas2015}, to cover each  line of source code, we need to make sure that each basic code block in a method is reached. In traditional search-based testing, this reachability would be expressed by a combination of branch distance \cite{McMinn:2004:SST:1077276.1077279} and approach-level. The approach-level measures how distant an individual execution and the target statement are in terms of the control dependencies. The branch distance estimates how distant  a predicate (a decision making point) is from evaluation to a desired target result. For example, given a predicate x==6 and an execution with value x = 4, the branch distance to the predicate valuing true would be |4-6|=2, whereas execution with value x=5 is closer to being true with a branch distance of |5-6|=1. Branch distance can be measured by applying a set of standard rules \cite{McMinn:2004:SST:1077276.1077279,Korel:1990:AST:101747.101755}.

In addition to test case generation, if reformation is a test suite to execute all statements then the approach level is not important, as all statements will be executed by the similar test suite. Hence, we only need to inspect the branch distances of all the branches that are related to the control dependencies of any of the statements in that class. There is a control dependency for some statements for each conditional statement in the code. It is required that the branch of the statement leading to the dependent code is executed. Hence, by executing all the tests in a test suite the line coverage fitness value  can be calculated. The minimum branch distances \textit{$d_{min}(b,Suite)$}  are calculated for each executed statement among all observed executions to every branch \textit{b} in the collection of control dependent branches \textit{$B_{CD}$}. Thus, the line coverage fitness function is defined as \cite{Rojas2015}:

$$f_{LC}(Suite)= v(|NCLs|- |CoveredLines|)
+\sum_{b\in B_{CD}}v(d_{min}(b,Suite))$$\\

 Where \textit{NCLs} are the set of all statements in the CUT, \textit{CoveredLines} are the total set of covered statements which are executed by each test case in the test suite, and \textit{v(x)} is a normalizing function in [0,1] (e.g. \textit{v(x)} = $\frac{x}{(x+1)}$) \cite{5477082}.

\subsubsection{Branch Coverage}
The idea of covering branches is well accepted in practice and implemented in popular tools,  even though the practical rationale of branch coverage may not always match the more theoretical interpretation of covering all edges of a program's control flow. Branch coverage is often defined as maximizing the number of branches of conditional statements that are executed by a test suite. Thus, a unit test suite is considered as satisfied if and only if its at least one test case satisfies the branch predicate to \textit{true} and at least one test case satisfies the branch predicate to \textit{false}.

The fitness value for the branch coverage is calculated based on a criteria which is how close a test suite is to covering all branches of the CUT. The fitness value of a test suite is calculated by executing all of its test cases, keeping trail of the branch distances $d(b,Suite)$ for each of the branch in the CUT. Then \cite{Rojas2015}:

$$f_{BC}(Suite)= \sum_{b\in B}v(d(b,Suite))$$

To optimize the branch coverage the following distance is calculated, where $d_{min} (b,Suite)$ is the minimal branch distance of branch b on all executions for the test suite \cite{Rojas2015}:

$\\\\ d(b,Suite)= \left\{  \begin{array}{rcl}
0 & \textnormal{if the branch has been covered,}\\
 & \\ v(d_{min}(b,Suite))& \textnormal{if the predicate has been }\\ &   \textnormal{executed at least twice,}\\
 &  \\ 1 &  \textnormal{otherwise,}
\end{array} \right.\\ $

Here it is needed to cover the \textit{true} and \textit{false} evaluation of a predicate, so that a predicate must be executed at least twice by a test suite. If the predicate is executed only once, then in theory the searching could oscillate between \textit{true} and \textit{false}.

\subsubsection{Weak Mutation}
Test case generation tools prefer to generate values that satisfy the constraints or conditions, rather than developers preferred values like  boundary cases. In weak mutation a small code modification is applied to the CUT and then force the test generation tool to generate such values that can distinguish between the original and the mutant. If the execution of a test case on the mutant leads to a different output than the execution on the CUT than a mutant is considered to be "killed" in the weak mutation. A test suite satisfies the weak mutation criterion if and only if at least one test case kill each mutant for the CUT.

Infection distance is measured with respect to a set of mutation operator which guides to calculate the fitness value for the weak mutation criterion. Here inference of a minimal infection distance function $d_{min}(\mu, Suite)$ exists and define \cite{Rojas2015}:

$$\\\\ d_w (\mu,Suite)= \left\{  \begin{array}{rcl}
1 & \textnormal{if mutant $\mu$ was not reached,}\\
 & \\ v(d_{min}(\mu,Suite))& \textnormal{if mutant $\mu$ was reached. }\\ 
\end{array} \right.\\\\ $$

This results in the following fitness function for weak mutation \cite{Rojas2015}:

$$f_{WM}(Suite)= \sum_{\mu \in M_c}d_w(\mu,Suite)$$

Where $M_c$ is the set of all mutants generated for the CUT.

\section{Evaluation Method}
We conducted a set of experiments to answer the following research questions:
\begin{itemize}
\item \textbf{{RQ1:}} Which source test case generation technique(s) is/are most effective for MT in terms of fault detection?
\item \textbf{{RQ2:}} Can the source test case generation techniques be combined to increase the fault finding effectiveness of MT?
\item \textbf{{RQ3:}} Does the fault detection effectiveness of an individual MR change with the source test generation method?
\item \textbf{{RQ4:}} How does the source test suite size differ for each source test generation technique?
\end{itemize}
\subsection{Code Corpus}
We built a code corpus containing 77 functions that take numerical inputs and produce numerical outputs . We obtained these functions from the following open source projects:
\begin{itemize}
\item \textbf{{The Colt Project\footnote{http://acs.lbl.gov/software/colt/}:}} A set of open source libraries written for high-performance scientific and technical computing in Java.

\item \textbf{{Apache Mahout\footnote{https://mahout.apache.org/}:}} A machine learning library written in Java.

\item \textbf{{Apache Commons Mathematics Library\footnote{http://commons.apache.org/proper/commons-math/}:}} A library of lightweight and self-contained  mathematics and statistics components written in the Java.
\end{itemize}

We list these functions in Table \ref{tab:fulltable}. Functions in the code corpus perform various calculations using sets of numbers such as calculating statistics (e.g. average, standard deviation and kurtosis), calculating distances (e.g. Manhattan and Tanimoto) and searching/sorting. Lines of code of these functions varied between 4 and 52, and the number of input parameters for each function varied between 1 and 4.
\subsection{METtester}
\begin{figure}[t]
  \includegraphics[scale=0.6]{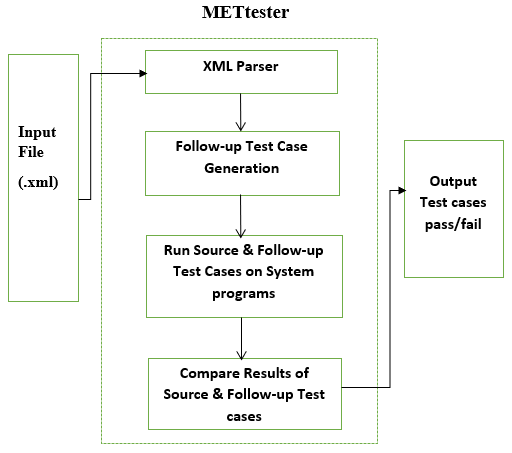}
  \caption{METtester Architecture.}
  \label{fig:2}
\end{figure}

METtester ~\cite{ps073006_2018_1157183} is a simple tool that we are developing to automate the MT process on a given Java program. This tool allows users to specify MRs and source test cases through a simple XML file. METtester transforms the source test cases according to the specified MRs and conducts MT on the given program. Figure~\ref{fig:2} shows the high level architecture of the tool.
Below we describe the important components of the tool:
\begin{itemize}
\item \textbf{{XML input file:}}  User will provide information (Figure \ref{fig:11}) regarding method names to test, source test inputs, MRs, and the number of test cases to run.
\begin{figure}[H]
  \includegraphics[scale=0.7]{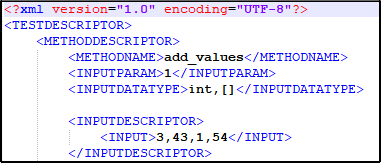}
  \caption{An example of the XML input given to METtester.}
  \label{fig:11}
\end{figure}
\item \textbf{{XML file parsing:}} Xmlparser class in our tool will parse information from the .xml file and process those. Then that information will be sent to the Follow-up test case generation module.
\item \textbf{{Follow-up test Case Generation:}} In this module follow-up test cases are generated based on the provided MRs and the source test cases.
\item \textbf{{Execute Source \& Follow-up test cases on the PUT:}}  After generation of the follow-up test cases METtester will run both the source and follow-up test cases individually into the system programs and return outputs from the programs.
\item \textbf{{Compare Source \& Follow-up test results:}} After getting the test results from the test program METtester will compare those results with the MR operators mentioned in the xml file. If it satisfies the MR property then the class will flag the test case as "Pass". If it fails to satisfy the MR property class will flag it as "Fail" which means there is fault in the program.
\end{itemize}
\subsection{Experimental Setup}
For the 77 methods described in Section 3.1 we generated a total of 7446 mutated versions using the $\mu$Java mutation tool \cite{Ma:2005:MAC:1077303.1077304}.  We used the following six metamorphic relations identified by Murphy et al. \cite{unknown} to test these functions:
\begin{itemize}
\item \textbf{{MR - Addition:}} Add a positive element. The expected result should be increased or remain constant.
\item \textbf{{MR - Multiplication:}} Multiply by a positive constant.  The expected result should be increased or remain constant.
\item \textbf{{MR - Shuffle:}} Randomly permute the elements. The expected result should remain constant.
\item \textbf{{MR - Inclusive:}} Add a new element. The expected result should increase or remain constant.
\item \textbf{{MR - Exclusive:}} Remove an existing element. The expected result should decrease or remain constant.
\item \textbf{{MR - Invertive:}} Take the inverse of each element. The expected result should decrease or remain constant.
\end{itemize}

For each of the methods, we used EvoSuite \cite{Fraser:2011:EAT:2025113.2025179} described in section $2.2$ to generate test cases targeting line, branch and weak mutation coverage. We used the generated test cases as the source test cases to conduct MT on the methods using the MRs described using METtester. Further, we randomly generated 10 test cases for each method to use as source test cases, to be used as the baseline.
\section{Results and Discussion}
\subsection{Effectiveness of the Source Test Case Generation Techniques }
\begin{figure}[H]
  \includegraphics[scale=0.5]{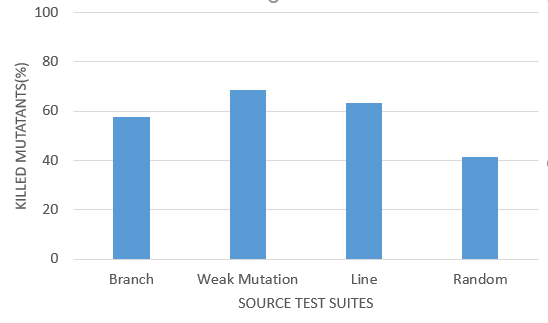}
  \caption{Total \% of mutants killed by each source test suite generation technique.}
  \label{fig:3}
\end{figure}

\begin{figure}[t]
  \includegraphics[scale=0.6]{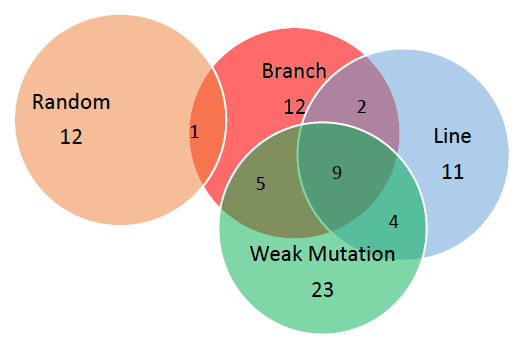}
  \caption{Venn Diagram for all the combinations of source test suites that performed best for each individual methods.}
  \label{fig:4}
\end{figure}

Figure \ref{fig:3} shows the overall mutant killing rates for the four source test generation techniques. Among all test case generation techniques, weak mutation performed best by killing 68.7\% mutants.  Random tests killed 41.5\% of the mutants.
Table \ref{tab:killrate} lists the number of methods that reported the highest mutant kill rates for each type of test generation technique. For some methods, several source test generation techniques gave the same best performance.Therefore, Figure \ref{fig:4} shows a Venn diagram of all the possible logical relations between the best performing source test generation techniques for the set of methods. Weak mutation based test generation technique reported the highest kill rate in 41 (53\%) methods, whereas random testing reported the highest kill rate only in 13 (17\%) methods. Therefore these results suggest that weak mutation based source test case generation is more effective in detecting faults with MT.
\begin{table}[H]
  \caption{Total number of methods having the highest mutants kill rate for each source test generation techniques.}
  \label{tab:killrate}
  \begin{tabular}{ccccc}
    \toprule
    Total Methods & Weak mutation & Line & Branch & Random\\
   \midrule
    77 & 41 & 26 & 29 & 13\\
    \bottomrule
  \end{tabular}
\end{table}
\fbox{\begin{minipage}{23em}
\textbf{RQ1:} Weak mutation based test suites have the highest fault detection rate for majority of the methods
\end{minipage}}
\\
\subsection{Fault Finding Effectiveness of  Combined Source Test Cases}

\begin{table*}[ht]
  \caption{All methods with Mutants kill rates and test suite size for each Source test case generation technique}
  \label{tab:fulltable}
 
  \resizebox{14.0cm}{!}{
  \begin{tabular}{|p{0.7\textwidth}|p{0.05\textwidth}|p{0.04\textwidth}|p{0.05\textwidth}|p{0.04\textwidth}|p{0.05\textwidth}|p{0.04\textwidth}|p{0.05\textwidth}|p{0.04\textwidth}|}
    \toprule  
    & \multicolumn{2}{c}{Branch} & \multicolumn{2}{c}{weak  mutation} & \multicolumn{2}{c}{Line} & \multicolumn{2}{c}{Random}|\\
    \toprule
    Method name & Killrate\newline(\%) & No. of Test Cases & Killrate\newline(\%) & No. of Test Cases  & Killrate\newline(\%) & No. of Test Cases & Killrate\newline(\%) & No. of Test Cases\\ 

   \midrule
   add\_values (Add elements in an array) & 63.63 & 1 & 63.63 &	1 & 54.54 & 1 & 30 & 10	\\
   
   array\_calc1&33.33&	1&33.33&	1&	46.15&1&52.10&	10	\\

array\_copy	(Deep copy an array)&56.00	&1	&64.00	&1	&64.00	&1	&0.00	&10\\

average	( Average of an array)&38.10&	1&	73.80&	1	&42.86&	1&	28.20&	10\\

bubble (Implements bubble sort)&	51.40&	1&	44.95&	3	&36.69	&1	&16.90&	10\\

cnt\_zeroes (Count zero in an array)&	41.00&	1&	51.30&	2&	38.46&	1	&0.00	&10\\

count\_k (Occurrences of k in an array)&	31.80&	1	&36.36&	2&	34.09&	1&	50.00&	10\\

count\_non\_zeroes (Count non zero element in array)&	41.00&1&48.71&	2&51.28&	1&22.20&10\\

dot\_product&	63.00&	1&	60.87&	1&	56.52&	1&	22.20&	10\\

elementwise\_max (Elementwise maximum)&	46.30&	2	&68.51&	3	&83.33&	2&	0.00&	10\\

elementwise\_min (Elementwise minimum)&	44.40&	1&	55.56&	1&	55.56&	1&	0.00&	10\\

find\_euc\_dist (Euclidean distance between two vectors)&	80.10&	1&	76.39&	1&	79.17&	1&	50&	10\\

find\_magnitude (Magnitude of a vector)&	52.10&	1&	75.00&	1	&52.10&	1&	8.69&	10\\

find\_max (find the maximum value)&	70.80&1	&50.00&	1	&50.00&	1	&70.90&	10\\
 
find\_max2&64.10&	1&	71.84&	2&	67.96&	1	&98.40&	10\\
 
find\_median (Find median value in an array)&	48.70	&2	&98.93	&3	&41.71	&2	&53.10	&10\\

find\_min (Find minimum value in an array)&	40.40&	1&	61.70	&1	&57.45&	1&	83.80&	10\\
 
geometric\_mean (Returns the geometric mean of the entries in the input array)&	51.20&	1	&53.66&	1&	95.12&	1&	65.40&	10\\

hamming\_dist (Hamming distance between two vectors)&	40.90&	1&	84.09&	3&	59.09&	2&	15.90&	10\\

insertion\_sort (Implements insertion sort)&	43.60&	1&42.55	&2&	37.23&	1&	32.65&	10\\

manhattan\_dist (Manhattan distance between two vectors)&	53.30	&1	&61.36&	2&	53.30&	1&	0.00&	10\\

mean\_absolute\_error (Measure of difference between two continuous variables)&	37.50&	1&	41.07&	2&	39.29	&1	&0.00	&10\\
 
selection\_sort (Implements selection sort)	&41.30&	1&	41.30&	2&	39.40&	1&	21.60&	10\\

sequential\_search (Finding a target value within a list)&	37.20&	2&	25.58&	3&	30.23&	2&	37.50&	10\\

set\_min\_val (Set array elements less than k equal to k)&	51.20&	2&	58.14&	2	&30.23&	1&	100&	10\\

shell\_sort (Implements shell sort)&	43.70&	1	&42.51	&1	&43.11	&1	&0.00&	10\\

variance (Returns the variance from a standard deviation)&	26.10&	1&	39.86&	1&	30.40&	1&	25.70&	10\\

weighted\_average (A mean calculated by giving values in a data set)	&86.10&	1&	56.94&	1&	86.10&	1	&21.20&	10\\
									
manhattanDistance (The distance between two points in a grid)&	48.89&	1&	77.78&	2&	22.22&	1&	9.10&	10	\\
 								
chebyshevDistance (Distance  metric defined on a vector space)&	39.08&	2&	43.68&	5&	35.63&	2&	2.00&	10	\\
								
tanimotoDistance (a proper distance metric)&	30.21&	2&	32.97&	5	&44.50&	2&	5.60&	10	\\
							
errorRate&	61.04&	3	&58.44&	2&	58.44&	2&	0.00&	10	\\
							
sum	&50.00&	1	&77.78&	1&	50.00&	1	&35.30&	10	\\
	
 distance1 (Compute the distance between the instance and another vector)&	53.33&	1&	80.00&	1&	53.33&	1&	14.8&	10\\
	
distanceInf (Compute the distance between the instance and another vector)&46.67&	1&	46.67&	1&	46.67&	1	&14.8&	10\\
	
ebeadd (Creates an array whose contents will be the element-by-element addition of the arguments)&	92.68&	2&	100.00&	3	&100.00&2&	15.8&	10\\
	
ebedivide (Creates an array whose contents will be the element-by-element division)&	100.00&	2&	100.00&	5	&100.00&	2&	26.8&	10\\

ebemultiply (Creates an array whose contents will be the element-by-element multiplication)&	100.00&	2&	100.00&	3&	92.68	&2&	15&	10\\
	
safeNorm (Returns the Cartesian norm )&	14.78&	1&	98.63&	5&97.08&	4&	0.8	&10\\
	
scale(Create a copy of an array scaled by a value)&	48.72&	1	&58.97&	3	&53.85	&1	&47.8&	10\\
	
entropy&	88.42&	1&	88.42&	2&	88.42&	1&	42.9&	10\\
	
g&	93.55&	2	&95.16	&2	&93.55&	1&	20.9&	10\\
	
calculateAbsoluteDifferences&	60.98&	1&	60.98&	1	&60.98	&1&	0&	10\\
	
evaluateHoners	&46.03&	1	&79.37&	1	&47.62&	1	&80.4&	10\\

evaluateInternal&	95.25&	1&	93.47&	2&	95.55&	1	&90.6	&10\\
	
evaluateNewton&	80.00&	1	&65.71	&1&64.29&	1&	76.8&	10\\
	
meanDifference (Returns the mean of the (signed) differences)&	40.00	&1	&80.00	&1	&40.00	&1	&40&	10\\
	
equals&	22.50&	3	&27.50	&4	&21.25	&3	&100&	10\\
 
chiSquare (Implements Chi-Square test statistics)&	96.41&	2	&96.41&	2	&96.41	&2	&65.6&	10\\
	
partition&	43.26&	5	&95.81&	5&	28.84&	3	&88.1&	10\\
	
evaluateWeightedProduct&	30.61&	2	&40.82&	2&	42.86&	2&	2&	10\\
	
autoCorrelation (Returns the auto-correlation of a data sequence)&	25.20&	2&	93.50&	2&	43.09&	1&	79.40&	10\\

covariance (Returns the covariance of two data sequences)&	24.84& 1&	23.57 &1&	23.57	&1&	86.70&	10\\
 
durbinWatson (Durbin-Watson computation)&	0.00	&0	&33.77	&1	&0.00	&0&	14.10&	10\\

harmonicMean (Returns the harmonic mean of a data sequence)&	74.00	&1	&74.00	&1	&76.00	&1	&42.50&	10\\

kurtosis (Returns the kurtosis (aka excess) of a data sequence)&	93.84	&1	&93.84	&1	&97.16	&1	&34.80&	10\\

lag1 (Returns the lag-1 autocorrelation of a dataset)&	99.55	&1	&32.70	&1	&89.55	&1	&33.70&	10\\

max (Returns the largest member of a data sequence)&	51.72	&1	&56.90	&1	&51.72	&1	&96.60&	10\\

meanDeviation (Returns the mean deviation of a dataset)	&54.39	&1	&33.33	&1	&28.07	&1	&78.30&	10\\
 
min (Returns the smallest member of a data sequence)&	67.41	&1	&81.03	&2	&70.69	&1	&96.60&	10\\
 
polevl &	94.23	&2	&88.46	&1	&88.46	&2	&45.50&	10\\

pooledMean (Returns the pooled mean of two data sequences)&	36.43	&1	&34.88&	1	&34.88	&1&19.30&	10\\
 
pooledVariance (Returns the pooled variance of two data sequences)&	43.08&	1	&47.83	&1	&47.83	&1	&31.10&	10\\

power &	53.33	&1	&53.33&	1&	53.33	&1	&15.80&	10\\

product	(Returns the product)&50.00&	1&	50.00	&1	&50.00&1&	94.70&	10\\

quantile (Returns the phi-quantile)&	40.13&	2&40.76&	2&	32.48&2	&40.00&10\\
 
sampleKurtosis ( Returns the sample kurtosis (aka excess) of a data sequence)&	93.86&	1&	93.86&1&	92.98&	1&	85.10&	10\\

sampleSkew (Returns the sample skew of a data sequence)&	89.47&	1	&89.47&	1	&97.37&	1&	89.50&	10\\

sampleVariance	(Returns the sample variance of a data sequence)&75.31&	1&	75.31&	1&	12.35&	1	&71.20&	10\\
 
skew ( Returns the skew of a data sequence)&	93.88&	1	&93.88&1	&93.88&	1&	48.80&	10\\

square &	47.37&	1&	47.37&	1	&57.89&	1&	5.30&	10\\

standardize (Modifies a data sequence to be standardized)&	89.26&	1&	89.26&	1	&91.95&	1&	77.60&10\\
 
sumOfLogarithms ( Returns the sum of logarithms of a data sequence)&	75.00&	1&	68.75&	1&	68.75&	1	&21.90&10\\

sumOfPowerOfDeviations&	68.75&	1&	52.08&	1&	75.00&	1&	64.90&	10\\

weightedMean (Returns the weighted mean of a data sequence)	&77.46&	1&	77.46&1&	77.46&	1	&65.00&	10\\

weightedRMS (Returns the weighted RMS (Root-Mean-Square) of a data sequence)&	86.96&	1&	86.96&	1&	86.96&	1&	43.30	&10\\

winsorizedMean (Returns the winsorized mean of a sorted data sequence)&	33.00&	1	&37.93&	1	&34.48	&1	&0.00&	10\\
    \bottomrule
  \end{tabular}}
\end{table*}
To observe whether combining source test case 
 generation techniques will achieve a higher fault detection rate, we combined the best performing source test generation technique, weak mutation, with the other source test generation techniques. 
 Table \ref{tab:combo} shows the total percentage of mutants killed with each combined test  suite. Combination of weak mutation and random test cases has the greater percentage of mutants kill rate (74.91) than combination of line (72.87) and branch (74.6) separately with weak mutation. If we combine all of the three strategies it slightly increases the total percentage of killed mutants (75.98)  but there are few things to be considered, like combined test suite size.
\begin{table}[H]
  \caption{Total \% of mutants killed after combining Weak Mutation, Line, Branch Coverage, and Random Testing }
  \label{tab:combo}
  \begin{tabular}{p{0.07\textwidth}p{0.1\textwidth}p{0.1\textwidth}p{0.07\textwidth}}
    \toprule
Weak Mutation +Line(\%)& Weak Mutation +Branch(\%) & Weak \newline Mutation\newline+Line+\newline Branch(\%)&Weak Mutation+\newline Random(\%)\\
    \midrule
    72.87 & 74.6 & 75.98 & 74.91\\
    \bottomrule
  \end{tabular}
\end{table}
\fbox{\begin{minipage}{23em}
 \textbf{RQ2:} Combining weak mutation test cases with random test cases will lead to detect more faults
\end{minipage}}
\\
\subsection{Fault Finding Effectiveness of Individual MRs}
\begin{figure*}[t]
  \centering
  \includegraphics[scale=0.8]{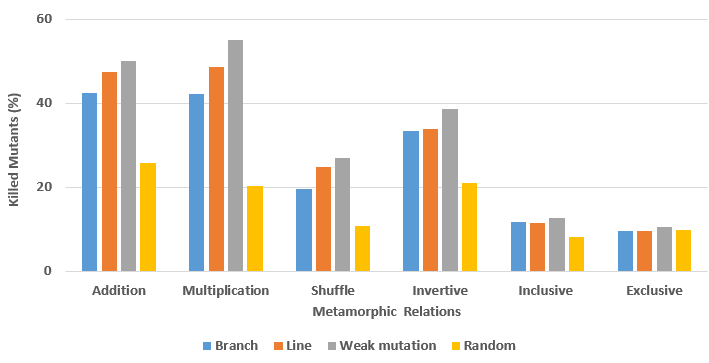}
\caption{\% of Mutants killed by all six MRs using 4 test suite strategies (Branch, Line Coverage, Weak Mutation and Random) }
\label{fig:fig}
\end{figure*}

 To see how each source test case generation technique performs with individual MRs, Figure \ref{fig:fig} illustrates the percentage of mutants killed by all six MRs separately using weak mutation, line, branch coverage and random test suites. Weak mutation has the highest percentage of killed mutants in all the six MRs. Specifically with multiplication and invertive MRs, the weak mutation test suite surpasses others on mutants' killing rate. But line coverage based test suites were similar  to weak mutation on  killing mutants with addition, shuffle, inclusive and exclusive MRs. For exclusive MR, all the test suites performed almost similarly.\\

\fbox{\begin{minipage}{23em}
\textbf{RQ3:} Weak mutation killed highest number of mutants in all the MRs
\end{minipage}}
\\
\subsection{Impact of Source Test Suite Size}
Table \ref{tab:averagetestsuite} compares the coverage criteria in terms of the total number of tests generated, their average and median test suite size of the individual methods. In addition, in columns Smaller, Equal, and Larger we compare whether the size of the weak mutation test suites are smaller, equal or larger than those produced by other source test case generation techniques. And p-value column shows the p-value computed using the paired t-test between weak mutation - line and weak mutation -branch. We are not comparing random test suites here, because we intentionally generated 10 random test cases for each method. Weak Mutation leads to larger test suites than branch and line coverage and on average, number of test cases produced for weak mutation are larger than those produced for branch and line coverage. The total number of test cases are also relatively larger for weak mutation compared to line and branch coverage.
\begin{table*}[t]
  \caption{Average test suites size for Weak mutation, Line coverage, Branch coverage and Random}
  \label{tab:averagetestsuite}
  \begin{tabular}{lccccccccc}
    \toprule
    Test Suites & Total Number of Test Cases & Average Size& Median size & Std Dev& Smaller & Equal & Larger&p-value\\
    \midrule
    Weak mutation & 135 & 1.75&1 & 1.13 & - & - & -&-\\
    Line & 97 & 1.26&1 & 0.67 & 1 & 45 & 31&3.102e-07\\
    branch & 99 & 1.29&1 & 0.59 & 2 & 49 & 26&1.375e-05\\
    Random & 770 & 10&10 & 0 & 77 & 0 & 0&-\\
    \bottomrule
  \end{tabular}
\end{table*}
\\

\fbox{\begin{minipage}{23em}
\textbf{RQ4:} Weak Mutation generated a higher number of test cases
\end{minipage}}
\section{Threats to validity}
Threats to \textit{internal validity} may result from the way empirical study was carried out. EvoSuite and our experimental setup have been carefully tested, although testing can not definitely prove the absence of defects.
 
Threats to \textit{construct validity} may occur because of the third party tools we have used. The
EvoSuite tool has been used to generate source test cases for line, branch and weak mutation test generation techniques. Further, we used the $\mu$Java mutation tool to create mutants for our experiment. To minimize these threats we verified that the results produced by these tools are correct by manually inspecting randomly selected outputs produced by each tool.
 
Threats to \textit{external validity} were minimized by using the 77 methods was employed as case study, which is collected from 4 different open source project classes. This provides high confidence in the possibility to generalize our results to other open source software. We only used the EvoSuite tool to generate test cases for our major experiment. But we also used the JCUTE \cite{conf/cav/SenA06} tool to generate branch coverage based test suites for our initial case study and also observed similar results.
\section{Related Work}
Most contributions on MT use either random generated test data or existing test suites for the  generation of source test cases. Not much research has been done on systematic generation of source test cases for MT. Gotlieb and Botella \cite{1245319} presented an approach called \textit{Automated Metamorphic Testing} where they translated the code into an equivalent constraint logic program and tried to find test cases that violates the MRs. Chen et al. \cite{DBLP:ChenKLT04} compared the effectiveness of random testing and "special values" as source test cases for MT. Special values are inputs where the output is well known for a particular method. Wu et al.\cite{article} proved that random test cases are  more effective than those test cases that are derived from "special values". Segura et al. \cite{SEGURA2011245} also  compared the effectiveness of random testing with manually generated test suites for MT. Their  results showed that randomly generated test suites are more effective in detecting faults than manually designed test suites. They also observed that combining random testing with manual tests provides better fault detection ability than random testing only.  

Batra and Sengupta \cite{10.1007/978-3-642-19423-8_19} proposed genetic algorithm to generate test cases maximizing the paths traversed in the program under test for MT. Chen et al. \cite{6528673} also addressed the same problem from a different prospective. They proposed partitioning the input domain of the PUT into multiple  equivalence classes for MT. They proposed an algorithm which will generate test cases which will cover those equivalence classes. They were able to generate test cases that provide high fault detection rate. Symbolic Execution was used to construct MRs and their corresponding source test cases by Dong and Zhang \cite{6615286}. Program paths were first analyzed to generate symbolic inputs and then, these symbolic inputs were used to construct MRs. In the final step, source test cases were generated by replacing the symbolic inputs with real values.
\section{Conclusions \& Future Work}
In this study we empirically evaluated the fault finding effectiveness of four different source test case generation strategies for MT: line, branch, weak mutation and random.

Our results show that weak mutation coverage based test generation can be an effective source test case generation technique for MT than the other techniques. Our results also show that the fault finding effectiveness of MT can be improved by combining source tests generated for weak mutation coverage with randomly generated source test cases. 

Further, in this paper we introduce a MT tool called "METtester." We plan to incorporate the investigated automated source test generation techniques into this tool. We also plan to extend the current case study to larger code bases and experiment with more source test generation techniques such as adaptive random test generation and data flow based test generation.  



\appendix


\begin{acks}
This work is supported by award number 1656877 from the National Science Foundation. Any Opinions, findings and conclusions or recommendations expressed in this material are those of the author(s) and do not necessarily reflect those of the National Science Foundation.


\end{acks}

\bibliographystyle{ACM-Reference-Format}
\bibliography{sample-bibliography}




 @String{AMSTrans = "American Mathematical Society Translations" }
 @String{AMSTrans = "Amer. Math. Soc. Transl." }
 @String{BullAMS = "Bulletin of the American Mathematical Society" }
 @String{BullAMS = "Bull. Amer. Math. Soc." }
 @String{ProcAMS = "Proceedings of the American Mathematical Society" }
 @String{ProcAMS = "Proc. Amer. Math. Soc." }
 @String{TransAMS = "Transactions of the American Mathematical Society" }
 @String{TransAMS = "Trans. Amer. Math. Soc." }

 @String{CACM = "Communications of the {ACM}" }
 @String{CACM = "Commun. {ACM}" }
 @String{CompServ = "Comput. Surveys" }
 @String{JACM = "J. ACM" }
 @String{ACMMathSoft = "{ACM} Transactions on Mathematical Software" }
 @String{ACMMathSoft = "{ACM} Trans. Math. Software" }
 @String{SIGNUM = "{ACM} {SIGNUM} Newsletter" }
 @String{SIGNUM = "{ACM} {SIGNUM} Newslett." }

 @String{AmerSocio = "American Journal of Sociology" }
 @String{AmerStatAssoc = "Journal of the American Statistical Association" }
 @String{AmerStatAssoc = "J. Amer. Statist. Assoc." }
 @String{ApplMathComp = "Applied Mathematics and Computation" }
 @String{ApplMathComp = "Appl. Math. Comput." }
 @String{AmerMathMonthly = "American Mathematical Monthly" }
 @String{AmerMathMonthly = "Amer. Math. Monthly" }
 @String{BIT = "{BIT}" }
 @String{BritStatPsych = "British Journal of Mathematical and Statistical
          Psychology" }
 @String{BritStatPsych = "Brit. J. Math. Statist. Psych." }
 @String{CanMathBull = "Canadian Mathematical Bulletin" }
 @String{CanMathBull = "Canad. Math. Bull." }
 @String{CompApplMath = "Journal of Computational and Applied Mathematics" }
 @String{CompApplMath = "J. Comput. Appl. Math." }
 @String{CompPhys = "Journal of Computational Physics" }
 @String{CompPhys = "J. Comput. Phys." }
 @String{CompStruct = "Computers and Structures" }
 @String{CompStruct = "Comput. \& Structures" }
 @String{CompJour = "The Computer Journal" }
 @String{CompJour = "Comput. J." }
 @String{CompSysSci = "Journal of Computer and System Sciences" }
 @String{CompSysSci = "J. Comput. System Sci." }
 @String{Computing = "Computing" }
 @String{ContempMath = "Contemporary Mathematics" }
 @String{ContempMath = "Contemp. Math." }
 @String{Crelle = "Crelle's Journal" }
 @String{GiornaleMath = "Giornale di Mathematiche" }
 @String{GiornaleMath = "Giorn. Mat." } 

 @String{Computer = "{IEEE} Computer" }
 @String{IEEETransComp = "{IEEE} Transactions on Computers" }
 @String{IEEETransComp = "{IEEE} Trans. Comput." }
 @String{IEEETransAC = "{IEEE} Transactions on Automatic Control" }
 @String{IEEETransAC = "{IEEE} Trans. Automat. Control" }
 @String{IEEESpec = "{IEEE} Spectrum" } 
 @String{ProcIEEE = "Proceedings of the {IEEE}" }
 @String{ProcIEEE = "Proc. {IEEE}" } 
 @String{IEEETransAeroElec = "{IEEE} Transactions on Aerospace and Electronic
     Systems" }
 @String{IEEETransAeroElec = "{IEEE} Trans. Aerospace Electron. Systems" }

 @String{IMANumerAna = "{IMA} Journal of Numerical Analysis" }
 @String{IMANumerAna = "{IMA} J. Numer. Anal." }
 @String{InfProcLet = "Information Processing Letters" }
 @String{InfProcLet = "Inform. Process. Lett." }
 @String{InstMathApp = "Journal of the Institute of Mathematics and
     its Applications" }
 @String{InstMathApp = "J. Inst. Math. Appl." }
 @String{IntControl = "International Journal of Control" }
 @String{IntControl = "Internat. J. Control" }
 @String{IntNumerEng = "International Journal for Numerical Methods in
     Engineering" }
 @String{IntNumerEng = "Internat. J. Numer. Methods Engrg." }
 @String{IntSuper = "International Journal of Supercomputing Applications" }
 @String{IntSuper = "Internat. J. Supercomputing Applic." } 
 @String{Kibernetika = "Kibernetika" }
 @String{JResNatBurStand = "Journal of Research of the National Bureau
     of Standards" }
 @String{JResNatBurStand = "J. Res. Nat. Bur. Standards" }
 @String{LinAlgApp = "Linear Algebra and its Applications" }
 @String{LinAlgApp = "Linear Algebra Appl." }
 @String{MathAnaAppl = "Journal of Mathematical Analysis and Applications" }
 @String{MathAnaAppl = "J. Math. Anal. Appl." }
 @String{MathAnnalen = "Mathematische Annalen" }
 @String{MathAnnalen = "Math. Ann." }
 @String{MathPhys = "Journal of Mathematical Physics" }
 @String{MathPhys = "J. Math. Phys." }
 @String{MathComp = "Mathematics of Computation" }
 @String{MathComp = "Math. Comp." }
 @String{MathScand = "Mathematica Scandinavica" }
 @String{MathScand = "Math. Scand." }
 @String{TablesAidsComp = "Mathematical Tables and Other Aids to Computation" }
 @String{TablesAidsComp = "Math. Tables Aids Comput." }
 @String{NumerMath = "Numerische Mathematik" }
 @String{NumerMath = "Numer. Math." }
 @String{PacificMath = "Pacific Journal of Mathematics" }
 @String{PacificMath = "Pacific J. Math." }
 @String{ParDistComp = "Journal of Parallel and Distributed Computing" }
 @String{ParDistComp = "J. Parallel and Distrib. Comput." } 
 @String{ParComputing = "Parallel Computing" }
 @String{ParComputing = "Parallel Comput." }
 @String{PhilMag = "Philosophical Magazine" }
 @String{PhilMag = "Philos. Mag." }
 @String{ProcNAS = "Proceedings of the National Academy of Sciences
                    of the USA" }
 @String{ProcNAS = "Proc. Nat. Acad. Sci. U. S. A." }
 @String{Psychometrika = "Psychometrika" }
 @String{QuartMath = "Quarterly Journal of Mathematics, Oxford, Series (2)" }
 @String{QuartMath = "Quart. J. Math. Oxford Ser. (2)" }
 @String{QuartApplMath = "Quarterly of Applied Mathematics" }
 @String{QuartApplMath = "Quart. Appl. Math." }
 @String{RevueInstStat = "Review of the International Statisical Institute" }
 @String{RevueInstStat = "Rev. Inst. Internat. Statist." }

 @String{JSIAM = "Journal of the Society for Industrial and Applied
     Mathematics" }
 @String{JSIAM = "J. Soc. Indust. Appl. Math." }
 @String{JSIAMB = "Journal of the Society for Industrial and Applied
     Mathematics, Series B, Numerical Analysis" }
 @String{JSIAMB = "J. Soc. Indust. Appl. Math. Ser. B Numer. Anal." }
 @String{SIAMAlgMeth = "{SIAM} Journal on Algebraic and Discrete Methods" }
 @String{SIAMAlgMeth = "{SIAM} J. Algebraic Discrete Methods" }
 @String{SIAMAppMath = "{SIAM} Journal on Applied Mathematics" }
 @String{SIAMAppMath = "{SIAM} J. Appl. Math." }
 @String{SIAMComp = "{SIAM} Journal on Computing" }
 @String{SIAMComp = "{SIAM} J. Comput." }
 @String{SIAMMatrix = "{SIAM} Journal on Matrix Analysis and Applications" }
 @String{SIAMMatrix = "{SIAM} J. Matrix Anal. Appl." }
 @String{SIAMNumAnal = "{SIAM} Journal on Numerical Analysis" }
 @String{SIAMNumAnal = "{SIAM} J. Numer. Anal." }
 @String{SIAMReview = "{SIAM} Review" }
 @String{SIAMReview = "{SIAM} Rev." }
 @String{SIAMSciStat = "{SIAM} Journal on Scientific and Statistical
     Computing" }
 @String{SIAMSciStat = "{SIAM} J. Sci. Statist. Comput." }

 @String{SoftPracExp = "Software Practice and Experience" }
 @String{SoftPracExp = "Software Prac. Experience" } 
 @String{StatScience = "Statistical Science" }
 @String{StatScience = "Statist. Sci." }
 @String{Techno = "Technometrics" }
 @String{USSRCompMathPhys = "{USSR} Computational Mathematics and Mathematical
     Physics" }
 @String{USSRCompMathPhys = "{U. S. S. R.} Comput. Math. and Math. Phys." }
 @String{VLSICompSys = "Journal of {VLSI} and Computer Systems" }
 @String{VLSICompSys = "J. {VLSI} Comput. Syst." }
 @String{ZAngewMathMech = "Zeitschrift fur Angewandte Mathematik und
     Mechanik" }
 @String{ZAngewMathMech = "Z. Angew. Math. Mech." }
 @String{ZAngewMathPhys = "Zeitschrift fur Angewandte Mathematik und Physik" }
 @String{ZAngewMathPhys = "Z. Angew. Math. Phys." }


 @String{Academic = "Academic Press" }
 @String{ACMPress = "{ACM} Press" }
 @String{AdamHilger = "Adam Hilger" }
 @String{AddisonWesley = "Addison-Wesley" }
 @String{AllynBacon = "Allyn and Bacon" }
 @String{AMS = "American Mathematical Society" }
 @String{Birkhauser = "Birkha{\"u}ser" }
 @String{CambridgePress = "Cambridge University Press" }
 @String{Chelsea = "Chelsea" }
 @String{ClaredonPress = "Claredon Press" }
 @String{DoverPub = "Dover Publications" }
 @String{Eyolles = "Eyolles" }
 @String{HoltRinehartWinston = "Holt, Rinehart and Winston" }
 @String{Interscience = "Interscience" }
 @String{JohnsHopkinsPress = "The Johns Hopkins University Press" }
 @String{JohnWileySons = "John Wiley and Sons" }
 @String{Macmillan = "Macmillan" }
 @String{MathWorks = "The Math Works Inc." }
 @String{McGrawHill = "McGraw-Hill" }
 @String{NatBurStd = "National Bureau of Standards" }
 @String{NorthHolland = "North-Holland" }
 @String{OxfordPress = "Oxford University Press" }  
 @String{PergamonPress = "Pergamon Press" }
 @String{PlenumPress = "Plenum Press" }
 @String{PrenticeHall = "Prentice-Hall" }
 @String{SIAMPub = "{SIAM} Publications" }
 @String{Springer = "Springer-Verlag" }
 @String{TexasPress = "University of Texas Press" }
 @String{VanNostrand = "Van Nostrand" }
 @String{WHFreeman = "W. H. Freeman and Co." }

@inproceedings{conf/cav/SenA06,
    author = "Sen, Koushik and Agha, Gul",
    editor = "Ball, Thomas and Jones, Robert B.",
    title = "CUTE and jCUTE: Concolic Unit Testing and Explicit Path
             Model-Checking Tools",
    booktitle = "CAV",
    crossref = "conf/cav/2006",
    ee = "http://dx.doi.org/10.1007/11817963_38",
    keywords = "formal methods, software engineering",
    pages = "419-423",
    year = "2006",
}

@inproceedings{Fraser:2011:EAT:2025113.2025179,
 author = {Fraser, Gordon and Arcuri, Andrea},
 title = {EvoSuite: Automatic Test Suite Generation for Object-oriented Software},
 booktitle = {Proceedings of the 19th ACM SIGSOFT Symposium and the 13th European Conference on Foundations of Software Engineering},
 series = {ESEC/FSE '11},
 year = {2011},
 isbn = {978-1-4503-0443-6},
 location = {Szeged, Hungary},
 pages = {416--419},
 numpages = {4},
 url = {http://doi.acm.org/10.1145/2025113.2025179},
 doi = {10.1145/2025113.2025179},
 acmid = {2025179},
 publisher = {ACM},
 address = {New York, NY, USA},
 keywords = {assertion generation, search based soft- ware testing, test case generation},
} 
@ARTICLE{6963470, 
author={E. T. Barr and M. Harman and P. McMinn and M. Shahbaz and S. Yoo}, 
journal={IEEE Transactions on Software Engineering}, 
title={The Oracle Problem in Software Testing: A Survey}, 
year={2015}, 
volume={41}, 
number={5}, 
pages={507-525}, 
keywords={formal specification;program testing;contract-driven development;domain specific information;informal oracle guidance;informal specifications;metamorphic testing;oracle automation;software testing practice;software testing research;test oracle information;test oracle problem;Automation;Licenses;Market research;Probabilistic logic;Reliability;Software testing;Automatic testing;Test oracle;Testing formalism;automatic testing;testing formalism}, 
doi={10.1109/TSE.2014.2372785}, 
ISSN={0098-5589}, 
month={May},}
@article{Weyukerarticle,
author = {Weyuker, Elaine},
year = {1982},
month = {11},
pages = {},
title = {On Testing Non-Testable Programs},
volume = {25},
booktitle = {Computer Journal}
}
@INPROCEEDINGS{5477082, 
author={A. Arcuri}, 
booktitle={2010 Third International Conference on Software Testing, Verification and Validation}, 
title={It Does Matter How You Normalise the Branch Distance in Search Based Software Testing}, 
year={2010}, 
volume={}, 
number={}, 
pages={205-214}, 
keywords={genetic algorithms;program testing;simulated annealing;tree searching;branch coverage;branch distance;control flow graph;genetic algorithm;heuristics;normalizing function;search algorithm;search based software testing;simulated annealing;structural criteria;test data generation;Application software;Automatic control;Data flow computing;Flow graphs;Genetic algorithms;Laboratories;Simulated annealing;Software algorithms;Software engineering;Software testing;Branch Distance;Genetic Algorithms;Search Based Software Testing;Simulated Annealing;Test Data Generation;Theory}, 
doi={10.1109/ICST.2010.17}, 
ISSN={2159-4848}, 
month={April},}
@INPROCEEDINGS{6615286, 
author={Guowei Dong and Tao Guo and Puhan Zhang}, 
booktitle={2013 IEEE 4th International Conference on Software Engineering and Service Science}, 
title={Security assurance with program path analysis and metamorphic testing}, 
year={2013}, 
volume={}, 
number={}, 
pages={193-197}, 
keywords={program diagnostics;program testing;security of data;information security;metamorphic testing;mission-critical software;oracle problem;path-combination-based MT method;program path analysis;program structure;security assurance;Magnetic resonance imaging;metamorphic testing;program path analysis;security assurance}, 
doi={10.1109/ICSESS.2013.6615286}, 
ISSN={2327-0586}, 
month={May},}
@INPROCEEDINGS{6528673, 
author={Leilei Chen and Lizhi Cai and Jiang Liu and Zhenyu Liu and Shiyan Wei and Pan Liu}, 
booktitle={2012 6th International Conference on New Trends in Information Science, Service Science and Data Mining (ISSDM2012)}, 
title={An optimized method for generating cases of metamorphic testing}, 
year={2012}, 
volume={}, 
number={}, 
pages={439-443}, 
keywords={equivalence classes;program testing;ECCEM criterion;TCR;equivalence-class coverage for every metamorphic relation;metamorphic testing;optimized method;oracle problem;software testing;test case rate of utilization;ECCEM criterion;measurement criterion;metamorphic testing;test case rate of utilization}, 
doi={}, 
ISSN={}, 
month={Oct},}
@InProceedings{10.1007/978-3-642-19423-8_19,
author="Batra, Gagandeep
and Sengupta, Jyotsna",
editor="Dua, Sumeet
and Sahni, Sartaj
and Goyal, D. P.",
title="An Efficient Metamorphic Testing Technique Using Genetic Algorithm",
booktitle="Information Intelligence, Systems, Technology and Management",
year="2011",
publisher="Springer Berlin Heidelberg",
address="Berlin, Heidelberg",
pages="180--188",
isbn="978-3-642-19423-8"
}


@article{SEGURA2011245,
title = "Automated metamorphic testing on the analyses of feature models",
journal = "Information and Software Technology",
volume = "53",
number = "3",
pages = "245 - 258",
year = "2011",
issn = "0950-5849",
doi = "https://doi.org/10.1016/j.infsof.2010.11.002",
url = "http://www.sciencedirect.com/science/article/pii/S0950584910001904",
author = "Sergio Segura and Robert M. Hierons and David Benavides and Antonio Ruiz-Cortés",
keywords = "Metamorphic testing, Test data generation, Mutation testing, Feature models, Automated analysis, Product lines"
}
@article{article,
author = {Wu, Peng and Xiao-Chun, SHI and Jiang-Jun, TANG and Hui-Min, LIN},
year = {2005},
month = {07},
pages = {},
title = {Metamorphic Testing and Special Case Testing: A Case Study},
volume = {16},
booktitle = {Journal of Software}
}
@inproceedings{DBLP:ChenKLT04,
  author    = {Tsong Yueh Chen and
               Fei{-}Ching Kuo and
               Ying Liu and
               Antony Tang},
  title     = {Metamorphic Testing and Testing with Special Values},
  booktitle = {4th {IEEE} International Workshop on Source Code Analysis and Manipulation
               {(SCAM} 2004), 15-16 September 2004, Chicago, IL, {USA}},
  pages     = {128--134},
  year      = {2004},
  timestamp = {Wed, 15 Apr 2015 18:38:11 +0200},
  biburl    = {http://dblp.org/rec/bib/conf/snpd/ChenKLT04},
  bibsource = {dblp computer science bibliography, http://dblp.org}
}
@INPROCEEDINGS{1245319, 
author={A. Gotlieb and B. Botella}, 
booktitle={Proceedings 27th Annual International Computer Software and Applications Conference. COMPAC 2003}, 
title={Automated metamorphic testing}, 
year={2003}, 
volume={}, 
number={}, 
pages={34-40}, 
keywords={constraint handling;program testing;program verification;software fault tolerance;software prototyping;INKA;automated metamorphic testing;automatic testing;constraint logic programming;fault-based testing;program testing;structural testing;test data generation;test data generator;Application software;Automatic testing;Computer applications;Data mining;Logic programming;Logic testing;Manuals;Prototypes;Software testing;System testing}, 
doi={10.1109/CMPSAC.2003.1245319}, 
ISSN={0730-3157}, 
month={Nov},}
@article{Ma:2005:MAC:1077303.1077304,
 author = {Ma, Yu-Seung and Offutt, Jeff and Kwon, Yong Rae},
 title = {MuJava: An Automated Class Mutation System: Research Articles},
 journal = {Softw. Test. Verif. Reliab.},
 issue_date = {June 2005},
 volume = {15},
 number = {2},
 month = jun,
 year = {2005},
 issn = {0960-0833},
 pages = {97--133},
 numpages = {37},
 url = {http://dx.doi.org/10.1002/stvr.v15:2},
 doi = {10.1002/stvr.v15:2},
 acmid = {1077304},
 publisher = {John Wiley and Sons Ltd.},
 address = {Chichester, UK},
 keywords = {mutation testing, object-oriented programs, software testing},
} 
@INPROCEEDINGS{unknown,
author = {Murphy, Christian and Kaiser, Gail and Hu, Lifeng and Wu, Leon},
year = {2008},
month = {01},
pages = {867-872},
title = {Properties of Machine Learning Applications for Use in Metamorphic Testing.}
}
@article{Korel:1990:AST:101747.101755,
 author = {Korel, B.},
 title = {Automated Software Test Data Generation},
 journal = {IEEE Trans. Softw. Eng.},
 issue_date = {August 1990},
 volume = {16},
 number = {8},
 month = aug,
 year = {1990},
 issn = {0098-5589},
 pages = {870--879},
 numpages = {10},
 url = {http://dx.doi.org/10.1109/32.57624},
 doi = {10.1109/32.57624},
 acmid = {101755},
 publisher = {IEEE Press},
 address = {Piscataway, NJ, USA},
 keywords = {array indexes, automated software test data generation, automatic programming, backtracking, data structures, dynamic data structures, dynamic data-flow analysis, function-minimization methods, function-minimization search algorithms, input variables, minimisation, pointers, program behavior, program execution flow, program testing, search problems.},
} 

@Inbook{Rojas2015,
author="Rojas, Jos{\'e} Miguel
and Campos, Jos{\'e}
and Vivanti, Mattia
and Fraser, Gordon
and Arcuri, Andrea",
editor="Barros, M{\'a}rcio
and Labiche, Yvan",
title="Combining Multiple Coverage Criteria in Search-Based Unit Test Generation",
bookTitle="Search-Based Software Engineering: 7th International Symposium, SSBSE 2015, Bergamo, Italy, September 5-7, 2015, Proceedings",
year="2015",
publisher="Springer International Publishing",
address="Cham",
pages="93--108",
isbn="978-3-319-22183-0",
doi="10.1007/978-3-319-22183-0_7",
url="https://doi.org/10.1007/978-3-319-22183-0_7"
}


@article{Fraser:2015:ASM:2780084.2780170,
 author = {Fraser, Gordon and Arcuri, Andrea},
 title = {Achieving Scalable Mutation-based Generation of Whole Test Suites},
 journal = {Empirical Softw. Engg.},
 issue_date = {June      2015},
 volume = {20},
 number = {3},
 month = jun,
 year = {2015},
 issn = {1382-3256},
 pages = {783--812},
 numpages = {30},
 url = {http://dx.doi.org/10.1007/s10664-013-9299-z},
 doi = {10.1007/s10664-013-9299-z},
 acmid = {2780170},
 publisher = {Kluwer Academic Publishers},
 address = {Hingham, MA, USA},
 keywords = {Mutation testing, Search-based testing, Test case generation, Testing classes, Unit testing},
} 
@article{McMinn:2004:SST:1077276.1077279,
 author = {McMinn, Phil},
 title = {Search-based Software Test Data Generation: A Survey: Research Articles},
 journal = {Softw. Test. Verif. Reliab.},
 issue_date = {June 2004},
 volume = {14},
 number = {2},
 month = jun,
 year = {2004},
 issn = {0960-0833},
 pages = {105--156},
 numpages = {52},
 url = {http://dx.doi.org/10.1002/stvr.v14:2},
 doi = {10.1002/stvr.v14:2},
 acmid = {1077279},
 publisher = {John Wiley and Sons Ltd.},
 address = {Chichester, UK},
 keywords = {automated software test data generation, evolutionary algorithms, evolutionary testing, metaheuristic search, search-based software engineering, simulated annealing},
} 
@inproceedings{Godefroid:2005:DDA:1065010.1065036,
 author = {Godefroid, Patrice and Klarlund, Nils and Sen, Koushik},
 title = {DART: Directed Automated Random Testing},
 booktitle = {Proceedings of the 2005 ACM SIGPLAN Conference on Programming Language Design and Implementation},
 series = {PLDI '05},
 year = {2005},
 isbn = {1-59593-056-6},
 location = {Chicago, IL, USA},
 pages = {213--223},
 numpages = {11},
 url = {http://doi.acm.org/10.1145/1065010.1065036},
 doi = {10.1145/1065010.1065036},
 acmid = {1065036},
 publisher = {ACM},
 address = {New York, NY, USA},
 keywords = {automated test generation, interfaces, program verification, random testing, software testing},
} 
@Article{CHEN201060,
  author   = {Tsong Yueh Chen and Fei-Ching Kuo and Robert G. Merkel and T.H. Tse},
  title    = {Adaptive Random Testing: The ART of test case diversity},
  journal  = {Journal of Systems and Software},
  year     = {2010},
  volume   = {83},
  number   = {1},
  pages    = {60 - 66},
  issn     = {0164-1212},
  note     = {SI: Top Scholars},
  doi      = {https://doi.org/10.1016/j.jss.2009.02.022},
  keywords = {Software testing, Random testing, Adaptive random testing, Adaptive random sequence, Failure-based testing, Failure pattern},
  url      = {http://www.sciencedirect.com/science/article/pii/S0164121209000405},
}
@inproceedings{Pacheco:2007:RFR:1297846.1297902,
 author = {Pacheco, Carlos and Ernst, Michael D.},
 title = {Randoop: Feedback-directed Random Testing for Java},
 booktitle = {Companion to the 22Nd ACM SIGPLAN Conference on Object-oriented Programming Systems and Applications Companion},
 series = {OOPSLA '07},
 year = {2007},
 isbn = {978-1-59593-865-7},
 location = {Montreal, Quebec, Canada},
 pages = {815--816},
 numpages = {2},
 url = {http://doi.acm.org/10.1145/1297846.1297902},
 doi = {10.1145/1297846.1297902},
 acmid = {1297902},
 publisher = {ACM},
 address = {New York, NY, USA},
 keywords = {Java, automatic test generation, random testing},
} 
@INPROCEEDINGS{6319263, 
author={T. Y. Chen and F. C. Kuo and D. Towey and Z. Q. Zhou}, 
booktitle={2012 12th International Conference on Quality Software}, 
title={Metamorphic Testing: Applications and Integration with Other Methods: Tutorial Synopsis}, 
year={2012}, 
volume={}, 
number={}, 
pages={285-288}, 
keywords={program testing;metamorphic testing;oracle problem;software testing;test case executions;Debugging;Flyback transformers;Production facilities;Software;Software testing;Wireless communication},
doi={10.1109/QSIC.2012.21}, 
ISSN={1550-6002}, 
month={Aug},}

@article{Howden:1976:RPA:1313320.1313531,
 author = {Howden, W. E.},
 title = {Reliability of the Path Analysis Testing Strategy},
 journal = {IEEE Trans. Softw. Eng.},
 issue_date = {May 1976},
 volume = {2},
 number = {3},
 month = may,
 year = {1976},
 issn = {0098-5589},
 pages = {208--215},
 numpages = {8},
 url = {http://dx.doi.org/10.1109/TSE.1976.233816},
 doi = {10.1109/TSE.1976.233816},
 acmid = {1313531},
 publisher = {IEEE Press},
 address = {Piscataway, NJ, USA},
 keywords = {Path analysis, program correctness, program testing, symbolic evaluation, symbolic evaluation, Path analysis, program correctness, program testing},
} 

@Book{Beizer:1990:STT:79060,
 author = {Beizer, Boris},
 title = {Software Testing Techniques (2Nd Ed.)},
 year = {1990},
 isbn = {0-442-20672-0},
 publisher = {Van Nostrand Reinhold Co.},
 address = {New York, NY, USA},
}
@InProceedings{1372139,
  author    = {T. Y. Chen and F. C. Kuo and T. H. Tse and Zhi Quan Zhou},
  title     = {Metamorphic testing and beyond},
  booktitle = {Eleventh Annual International Workshop on Software Technology and Engineering Practice},
  year      = {2003},
  pages     = {94-100},
  month     = {Sept},
  doi       = {10.1109/STEP.2003.18},
  keywords  = {program debugging;program testing;follow-up test cases;metamorphic testing;program debugging;program proving;program testing;successful test case;test case selection strategy;testing oracle;Arithmetic;Australia Council;Computer science;Cryptography;Debugging;Humans;Information systems;Information technology;Software testing;System testing;Follow-up test cases;metamorphic testing;semi-proving;successful test case;test case selection strategy;testing oracle},
}
@inproceedings{Chen2004CaseSO,
  title={Case Studies on the Selection of Useful Relations in Metamorphic Testing},
  author={T. Y. Chen and De Hao Huang and T. H. Tse and Zhi Quan Zhou},
  year={2004},
}
@article{particle,
author = {P. Pargas, Roy and Harrold, Mary and R. Peck, Robert},
year = {2000},
month = {02},
pages = {},
title = {Test-Data Generation Using Genetic Algorithms},
volume = {9},
booktitle = {Software Testing, Verification and Reliability}
}
@inproceedings{Chen:2015:MTS:2819261.2819278,
 author = {Chen, Tsong Yueh},
 title = {Metamorphic Testing: A Simple Method for Alleviating the Test Oracle Problem},
 booktitle = {Proceedings of the 10th International Workshop on Automation of Software Test},
 series = {AST '15},
 year = {2015},
 location = {Florence, Italy},
 pages = {53--54},
 numpages = {2},
 url = {http://dl.acm.org/citation.cfm?id=2819261.2819278},
 acmid = {2819278},
 publisher = {IEEE Press},
 address = {Piscataway, NJ, USA},
 keywords = {metamorphic testing, test oracles},
} 

@Article{Abril07,
  author 	= "Patricia S. Abril and Robert Plant",
  title 	= "The patent holder's dilemma: Buy, sell, or troll?",
  journal 	= "Communications of the ACM",
  volume 	= "50",
  number 	= "1",
  month 	= jan,
  year 		= "2007",
  pages 	= "36--44",
  doi 		= "10.1145/1188913.1188915",
  url		= "http://doi.acm.org/10.1145/1219092.1219093",
  note		= "",
}

@Article{Cohen07,
  author 	=	"Sarah Cohen and Werner Nutt and Yehoshua Sagic",
  title		=	"Deciding equivalances among conjunctive aggregate queries",
  journal 	=	JACM,
  articleno	=	"5",
  numpages	=	"50",
  volume 	=	"54",
  number 	= 	"2",
  month 	=	apr,
  year 		=	"2007",
  doi 		=	"10.1145/1219092.1219093",
  url 		=	"http://doi.acm.org/10.1145/1219092.1219093",
  acmid		=	"1219093",
  note 		= 	"",
}


@periodical{JCohen96,
  key = 	 "Cohen",
  editor =       "Jacques Cohen",
  title =        "Special issue: Digital Libraries",
  journal =      CACM,
  volume =       "39",
  number = 	 "11",
  month =	 nov,
  year = 	 "1996",
}


@Book{Kosiur01,
  author =       "David Kosiur",
  title =        "Understanding Policy-Based Networking",
  publisher =    "Wiley",
  year =         "2001",
  address =      "New York, NY",
  edition =      "2nd.",
  editor = 	 "",
  volume = 	 "",
  number = 	 "",
  series = 	 "",
  month = 	 "",
  note = 	 "",
}


@Book{Harel79,
  author =       "David Harel",
  year =         "1979",
  title =        "First-Order Dynamic Logic",
  series =       "Lecture Notes in Computer Science",
  volume =       "68",
  address =      "New York, NY",
  publisher =    "Springer-Verlag",
  doi = 	 "10.1007/3-540-09237-4",
  url = 	 "http://dx.doi.org/10.1007/3-540-09237-4",
  editor = 	 "",
  number = 	 "",
  month = 	 "",
  note = 	 "",
}


@Inbook{Editor00,
  author = 	 "",
  editor =       "Ian Editor",
  title =        "The title of book one",
  subtitle =     "The book subtitle",
  series =       "The name of the series one",
  year =         "2007",
  volume = 	 "9",
  address =      "Chicago",
  edition =      "1st.",
  publisher =    "University of Chicago Press",
  doi = 	 "10.1007/3-540-09237-4",
  url = 	 "http://dx.doi.org/10.1007/3-540-09456-9",
  chapter = 	 "",
  pages = 	 "",
  number = 	 "",
  type = 	 "",
  month = 	 "",
  note = 	 "",
}

%
@InBook{Editor00a,
  author = 	 "",
  editor =       "Ian Editor",
  title =        "The title of book two",
  subtitle =     "The book subtitle",
  series =       "The name of the series two",
  year =         "2008",
  address =      "Chicago",
  edition =      "2nd.",
  publisher =    "University of Chicago Press",
  doi = 	 "10.1007/3-540-09237-4",
  url = 	 "http://dx.doi.org/10.1007/3-540-09456-9",
  volume = 	 "",
  chapter = 	 "100",
  pages = 	 "",
  number = 	 "",
  type = 	 "",
  month = 	 "",
  note = 	 "",
}


@Incollection{Spector90,
  author =       "Asad Z. Spector",
  title =        "Achieving application requirements",
  booktitle =    "Distributed Systems",
  publisher =    "ACM Press",
  address =      "New York, NY",
  year =         "1990",
  edition =      "2nd.",
  chapter =      "",
  editor = 	 "Sape Mullender",
  pages =        "19--33",
  doi = 	 "10.1145/90417.90738",
  url = 	 "http://doi.acm.org/10.1145/90417.90738",
  volume =	 "",
  number = 	 "",
  series =	 "",
  type = 	 "",
  month = 	 "",
  note = 	 "",
}


@Incollection{Douglass98,
  author =       "Bruce P. Douglass and David Harel and Mark B. Trakhtenbrot",
  title =        "Statecarts in use: structured analysis and object-orientation",
  series =       "Lecture Notes in Computer Science",
  booktitle =    "Lectures on Embedded Systems",
  publisher =    "Springer-Verlag",
  address =      "London",
  volume = 	 "1494",
  year =         "1998",
  chapter =      "",
  editor = 	 "Grzegorz Rozenberg and Frits W. Vaandrager",
  pages =        "368--394",
  doi = 	 "10.1007/3-540-65193-4_29",
  url = 	 "http://dx.doi.org/10.1007/3-540-65193-4_29",
  edition = 	 "",
  number = 	 "",
  type = 	 "",
  month = 	 "",
  note = 	 "",
}


@Book{Knuth97,
  author =       "Donald E. Knuth",
  title =        "The Art of Computer Programming, Vol. 1: Fundamental Algorithms (3rd. ed.)",
  publisher =    "Addison Wesley Longman Publishing Co., Inc.",
  year =         "1997",
  address =      "",
  edition =      "",
  editor = 	 "",
  volume = 	 "",
  number = 	 "",
  series = 	 "",
  month = 	 "",
  note = 	 "",
}


@Book{Knuth98,
  author =       "Donald E. Knuth",
  year =         "1998",
  title =        "The Art of Computer Programming",
  series =       "Fundamental Algorithms",
  volume =       "1",
  edition =      "3rd",
  address =      "",
  publisher =    "Addison Wesley Longman Publishing Co., Inc.",
  doi = 	 "",
  url = 	 "",
  editor = 	 "",
  number = 	 "",
  month = 	 "",
  note = 	 "(book)",
}






@incollection{GM05,
Author= "Dan Geiger and Christopher Meek",
Title= "Structured Variational Inference Procedures and their Realizations (as incol)",
Year= 2005,
Booktitle="Proceedings of Tenth International Workshop on Artificial Intelligence and Statistics, {\rm The Barbados}",
Publisher="The Society for Artificial Intelligence and Statistics",
Month= jan,
Editors= "Z. Ghahramani and R. Cowell"
}

@Inproceedings{Smith10,
  author =       "Stan W. Smith",
  title =        "An experiment in bibliographic mark-up: Parsing metadata for XML export",
  booktitle =    "Proceedings of the 3rd. annual workshop on Librarians and Computers",
  series = 	 "LAC '10",
  editor = 	 "Reginald N. Smythe and Alexander Noble",
  volume = 	  "3",
  year =         "2010",
  publisher =    "Paparazzi Press",
  address = 	 "Milan Italy",
  pages =        "422--431",
  doi = 	 "99.9999/woot07-S422",
  url = 	 "http://dx.doi.org/99.0000/woot07-S422",
  number =	 "",
  month = 	 "",
  organization = "",
  note = 	 "",
}

@Inproceedings{VanGundy07,
  author =       "Matthew Van Gundy and Davide Balzarotti and Giovanni Vigna",
  year =         "2007",
  title =        "Catch me, if you can: Evading network signatures with web-based polymorphic worms",
  booktitle =    "Proceedings of the first USENIX workshop on Offensive Technologies",
  series = 	 "WOOT '07",
  publisher =    "USENIX Association",
  address = 	 "Berkley, CA",
  articleno = 	 "7",
  numpages = 	 "9",
  editor = 	 "",
  volume = 	 "",
  number = 	 "",
  pages = 	 "",
  month = 	 "",
  organization = "",
  note = 	 "",
}

@Inproceedings{VanGundy08,
  author =       "Matthew Van Gundy and Davide Balzarotti and Giovanni Vigna",
  year =         "2008",
  title =        "Catch me, if you can: Evading network signatures with web-based polymorphic worms",
  booktitle =    "Proceedings of the first USENIX workshop on Offensive Technologies",
  series = 	 "WOOT '08",
  publisher =    "USENIX Association",
  address = 	 "Berkley, CA",
  articleno = 	 "7",
  numpages = 	 "2",
  editor = 	 "",
  volume = 	 "",
  number = 	 "",
  pages = 	 "99-100",
  month = 	 "",
  organization = "",
  note = 	 "",
}

@Inproceedings{VanGundy09,
  author =       "Matthew Van Gundy and Davide Balzarotti and Giovanni Vigna",
  year =         "2009",
  title =        "Catch me, if you can: Evading network signatures with web-based polymorphic worms",
  booktitle =    "Proceedings of the first USENIX workshop on Offensive Technologies",
  series = 	 "WOOT '09",
  publisher =    "USENIX Association",
  address = 	 "Berkley, CA",
  articleno = 	 "",
  numpages = 	 "",
  editor = 	 "",
  volume = 	 "",
  number = 	 "",
  pages = 	 "90--100",
  month = 	 "",
  organization = "",
  note = 	 "",
}

@Inproceedings{Andler79,
  author =       "Sten Andler",
  title =        "Predicate Path expressions",
  booktitle =    "Proceedings of the 6th. ACM SIGACT-SIGPLAN symposium on Principles of Programming Languages",
  series = 	 "POPL '79",
  year =         "1979",
  publisher =    "ACM Press",
  address = 	 "New York, NY",
  pages =        "226--236",
  doi = 	 "10.1145/567752.567774",
  url = 	 "http://doi.acm.org/10.1145/567752.567774",
  editor = 	 "",
  volume = 	 "",
  number = 	 "",
  month = 	 "",
  organization = "",
  note = 	 "",
}

@Techreport{Harel78,
  author =       "David Harel",
  year =         "1978",
  title =        "LOGICS of Programs: AXIOMATICS and DESCRIPTIVE POWER",
  institution =  "Massachusetts Institute of Technology",
  type =         "MIT Research Lab Technical Report",
  number =       "TR-200",
  address =      "Cambridge, MA",
  month = 	 "",
  note = 	 "",
}

@MASTERSTHESIS{anisi03,
author = {David A. Anisi},
title = {Optimal Motion Control of a Ground Vehicle},
school = {Royal Institute of Technology (KTH), Stockholm, Sweden},
intitution = {FOI-R-0961-SE, Swedish Defence Research Agency (FOI)},
year = {2003},
}


@Phdthesis{Clarkson85,
  author =       "Kenneth L. Clarkson",
  year =         "1985",
  title =        "Algorithms for Closest-Point Problems (Computational Geometry)",
  school =       "Stanford University",
  address =      "Palo Alto, CA",
  note =         "UMI Order Number: AAT 8506171",
  type = 	 "",
  month = 	 "",
}


@online{Thornburg01,
  author =       "Harry Thornburg",
  year =         "2001",
  title =        "Introduction to Bayesian Statistics",
  url =           "http://ccrma.stanford.edu/~jos/bayes/bayes.html",
  month = 	 mar,
  lastaccessed = "March 2, 2005",
}


@online{Ablamowicz07,
  author =       "Rafal Ablamowicz and Bertfried Fauser",
  year =         "2007",
  title =        "CLIFFORD: a Maple 11 Package for Clifford Algebra Computations, version 11",
  url =          "http://math.tntech.edu/rafal/cliff11/index.html",
  lastaccessed = "February 28, 2008",
}


@misc{Poker06,
  author =       "Poker-Edge.Com",
  year =         "2006",
  month 	= mar,
  title =        "Stats and Analysis",
  lastaccessed = "June 7, 2006",
  url =          "http://www.poker-edge.com/stats.php",
}

@misc{Obama08,
  author 	= "Barack Obama",
  year 		= "2008",
  title		= "A more perfect union",
  howpublished  = "Video",
  day 		= "5",
  url 		= "http://video.google.com/videoplay?docid=6528042696351994555",
  month 	= mar,
  lastaccessed  = "March 21, 2008",
  note		=  "",
}

@misc{JoeScientist001,
  author =       "Joseph Scientist",
  year =         "2009",
  title =        "The fountain of youth",
  note =         "Patent No. 12345, Filed July 1st., 2008, Issued Aug. 9th., 2009",
  url =           "",
  howpublished = "",
  month = 	 aug,
  lastaccessed = "",
}


@Inproceedings{Novak03,
  author =       "Dave Novak",
  title =        "Solder man",
  booktitle =    "ACM SIGGRAPH 2003 Video Review on Animation theater Program: Part I - Vol. 145 (July 27--27, 2003)",
  year =         "2003",
  publisher = 	 "ACM Press",
  address = 	 "New York, NY",
  pages =        "4",
  month = 	 "March 21, 2008",
  doi = 	 "99.9999/woot07-S422",
  url = 	 "http://video.google.com/videoplay?docid=6528042696351994555",
  note = 	 "",
  howpublished = "Video",
  editor = 	 "",
  volume = 	 "",
  number = 	 "",
  series = 	 "",
  organization = "",
}


@article{Lee05,
  author =       "Newton Lee",
  year = 	 "2005",
  title =        "Interview with Bill Kinder: January 13, 2005",
  journal = 	 "Comput. Entertain.",
  eid      =     "4",
  volume =       "3",
  number = 	 "1",
  month =  	 "Jan.-March",
  doi = 	 "10.1145/1057270.1057278",
  url = 	 "http://doi.acm.org/10.1145/1057270.1057278",
  howpublished = "Video",
  note =	 "",
}

@article{Rous08,
  author =       "Bernard Rous",
  year = 	 "2008",
  title =        "The Enabling of Digital Libraries",
  journal = 	 "Digital Libraries",
  volume =       "12",
  number = 	 "3",
  month =  	 jul,
  articleno = 	 "5",
  doi = 	 "",
  url = 	 "",
  howpublished = "",
  note =	 "To appear",
}

@article{384253,
 author = {Werneck,, Renato and Setubal,, Jo\~{a}o and da Conceic\~{a}o,, Arlindo},
 title = {(old) Finding minimum congestion spanning trees},
 journal = {J. Exp. Algorithmics},
 volume = {5},
 year = {2000},
 issn = {1084-6654},
 pages = {11},
 doi = {http://doi.acm.org/10.1145/351827.384253},
 publisher = {ACM},
 address = {New York, NY, USA},
 }


@article{Werneck:2000:FMC:351827.384253,
 author = {Werneck, Renato and Setubal, Jo\~{a}o and da Conceic\~{a}o, Arlindo},
 title = {(new) Finding minimum congestion spanning trees},
 journal = {J. Exp. Algorithmics},
 volume = {5},
 month = dec,
 year = {2000},
 issn = {1084-6654},
 articleno = {11},
 url = {http://portal.acm.org/citation.cfm?id=351827.384253},
 doi = {10.1145/351827.384253},
 acmid = {384253},
 publisher = {ACM},
 address = {New York, NY, USA},
}

@article{1555162,
 author = {Conti, Mauro and Di Pietro, Roberto and Mancini, Luigi V. and Mei, Alessandro},
 title = {(old) Distributed data source verification in wireless sensor networks},
 journal = {Inf. Fusion},
 volume = {10},
 number = {4},
 year = {2009},
 issn = {1566-2535},
 pages = {342--353},
 doi = {http://dx.doi.org/10.1016/j.inffus.2009.01.002},
 publisher = {Elsevier Science Publishers B. V.},
 address = {Amsterdam, The Netherlands, The Netherlands},
 }

@article{Conti:2009:DDS:1555009.1555162,
 author = {Conti, Mauro and Di Pietro, Roberto and Mancini, Luigi V. and Mei, Alessandro},
 title = {(new) Distributed data source verification in wireless sensor networks},
 journal = {Inf. Fusion},
 volume = {10},
 number = {4},
 month = oct,
 year = {2009},
 issn = {1566-2535},
 pages = {342--353},
 numpages = {12},
 url = {http://portal.acm.org/citation.cfm?id=1555009.1555162},
 doi = {10.1016/j.inffus.2009.01.002},
 acmid = {1555162},
 publisher = {Elsevier Science Publishers B. V.},
 address = {Amsterdam, The Netherlands, The Netherlands},
 keywords = {Clone detection, Distributed protocol, Securing data fusion, Wireless sensor networks},
}

@inproceedings{Li:2008:PUC:1358628.1358946,
 author = {Li, Cheng-Lun and Buyuktur, Ayse G. and Hutchful, David K. and Sant, Natasha B. and Nainwal, Satyendra K.},
 title = {Portalis: using competitive online interactions to support aid initiatives for the homeless},
 booktitle = {CHI '08 extended abstracts on Human factors in computing systems},
 year = {2008},
 isbn = {978-1-60558-012-X},
 location = {Florence, Italy},
 pages = {3873--3878},
 numpages = {6},
 url = {http://portal.acm.org/citation.cfm?id=1358628.1358946},
 doi = {10.1145/1358628.1358946},
 acmid = {1358946},
 publisher = {ACM},
 address = {New York, NY, USA},
 keywords = {cscw, distributed knowledge acquisition, incentive design, online games, recommender systems, reputation systems, user studies, virtual community},
}

@book{Hollis:1999:VBD:519964,
 author = {Hollis, Billy S.},
 title = {Visual Basic 6: Design, Specification, and Objects with Other},
 year = {1999},
 isbn = {0130850845},
 edition = {1st},
 publisher = {Prentice Hall PTR},
 address = {Upper Saddle River, NJ, USA},
 }


@book{Goossens:1999:LWC:553897,
 author = {Goossens, Michel and Rahtz, S. P. and Moore, Ross and Sutor, Robert S.},
 title = {The  Latex Web Companion: Integrating TEX, HTML, and XML},
 year = {1999},
 isbn = {0201433117},
 edition = {1st},
 publisher = {Addison-Wesley Longman Publishing Co., Inc.},
 address = {Boston, MA, USA},
 }


@techreport{897367,
 author = {Buss, Jonathan F. and Rosenberg, Arnold L. and Knott, Judson D.},
 title = {Vertex Types in Book-Embeddings},
 year = {1987},
 source = {http://www.ncstrl.org:8900/ncstrl/servlet/search?formname=detail\&id=oai
 publisher = {University of Massachusetts},
 address = {Amherst, MA, USA},
 }

@techreport{Buss:1987:VTB:897367,
 author = {Buss, Jonathan F. and Rosenberg, Arnold L. and Knott, Judson D.},
 title = {Vertex Types in Book-Embeddings},
 year = {1987},
 source = {http://www.ncstrl.org:8900/ncstrl/servlet/search?formname=detail\&id=oai
 publisher = {University of Massachusetts},
 address = {Amherst, MA, USA},
 }


@proceedings{Czerwinski:2008:1358628,
 author = {},
 note = {General Chair-Czerwinski, Mary and General Chair-Lund, Arnie and Program Chair-Tan, Desney},
 title = {CHI '08: CHI '08 extended abstracts on Human factors in computing systems},
 year = {2008},
 isbn = {978-1-60558-012-X},
 location = {Florence, Italy},
 order_no = {608085},
 publisher = {ACM},
 address = {New York, NY, USA},
 }


@phdthesis{Clarkson:1985:ACP:911891,
 author = {Clarkson, Kenneth Lee},
 advisor = {Yao, Andrew C.},
 title = {Algorithms for Closest-Point Problems (Computational Geometry)},
 year = {1985},
 note = {AAT 8506171},
 school = {Stanford University},
 address = {Stanford, CA, USA},
 }



@Article{1984:1040142,
 key = {{$\!\!$}},
 journal = {SIGCOMM Comput. Commun. Rev.},
 year = {1984},
 issn = {0146-4833},
 volume = {13-14},
 number = {5-1},
 issue_date = {January/April 1984},
 publisher = {ACM},
 address = {New York, NY, USA},
 }


@inproceedings{2004:ITE:1009386.1010128,
 key = {IEEE},
 title = {IEEE TCSC Executive Committee},
 booktitle = {Proceedings of the IEEE International Conference on Web Services},
 series = {ICWS '04},
 year = {2004},
 isbn = {0-7695-2167-3},
 pages = {21--22},
 url = {http://dx.doi.org/10.1109/ICWS.2004.64},
 doi = {http://dx.doi.org/10.1109/ICWS.2004.64},
 acmid = {1010128},
 publisher = {IEEE Computer Society},
 address = {Washington, DC, USA},
}

@book{Mullender:1993:DS(:302430,
 editor = {Mullender, Sape},
 title = {Distributed systems (2nd Ed.)},
 year = {1993},
 isbn = {0-201-62427-3},
 publisher = {ACM Press/Addison-Wesley Publishing Co.},
 address = {New York, NY, USA},
 }


@techreport{Petrie:1986:NAD:899644,
 author = {Petrie, Charles J.},
 title = {New Algorithms for Dependency-Directed Backtracking (Master's thesis)},
 year = {1986},
 source = {http://www.ncstrl.org:8900/ncstrl/servlet/search?formname=detail\&id=oai
 publisher = {University of Texas at Austin},
 address = {Austin, TX, USA},
 }

@MASTERSTHESIS{Petrie:1986:NAD:12345,
 author = {Petrie, Charles J.},
 title = {New Algorithms for Dependency-Directed Backtracking (Master's thesis)},
 year = {1986},
 source = {http://www.ncstrl.org:8900/ncstrl/servlet/search?formname=detail\&id=oai
 school = {University of Texas at Austin},
 address = {Austin, TX, USA},
 }




@BOOK{book-minimal,
   author = "Donald E. Knuth",
   title = "Seminumerical Algorithms",
   publisher = "Addison-Wesley",
   year = "1981",
}

@INcollection{KA:2001,
 author = {Kong, Wei-Chang},
 Title = {The implementation of electronic commerce in SMEs in Singapore (as Incoll)},
 booktitle = {E-commerce and cultural values},
 year = {2001},
 isbn = {1-59140-056-2},
 pages = {51--74},
 numpages = {24},
 url = {http://portal.acm.org/citation.cfm?id=887006.887010},
 acmid = {887010},
 publisher = {IGI Publishing},
 address = {Hershey, PA, USA},
}


@INBOOK{KAGM:2001,
 author = {Kong, Wei-Chang},
 type = {Name of Chapter:},
 chapter = {The implementation of electronic commerce in SMEs in Singapore (Inbook-w-chap-w-type)},
 title = {E-commerce and cultural values},
 year = {2001},
 isbn = {1-59140-056-2},
 pages = {51--74},
 numpages = {24},
 url = {http://portal.acm.org/citation.cfm?id=887006.887010},
 acmid = {887010},
 publisher = {IGI Publishing},
 address = {Hershey, PA, USA},
}




@incollection{Kong:2002:IEC:887006.887010,
  author = 	{Kong, Wei-Chang},
  editor =	{Theerasak Thanasankit},
  title =	{Chapter 9},
  booktitle =	{E-commerce and cultural values (Incoll-w-text (chap 9) 'title')},
  year =	{2002},
  address =	{Hershey, PA, USA},
  publisher =	{IGI Publishing},
  url =		{http://portal.acm.org/citation.cfm?id=887006.887010},
  pages =	{51--74},
  numpages =	{24},
  acmid =	{887010},
  isbn =	{1-59140-056-2},
  number = 	 "",
  type = 	 "",
  month = 	 "",
  note = 	 "",
}

@incollection{Kong:2003:IEC:887006.887011,
 author = {Kong, Wei-Chang},
 title = {The implementation of electronic commerce in SMEs in Singapore (Incoll)},
 booktitle = {E-commerce and cultural values},
 editor = {Thanasankit, Theerasak},
 year = {2003},
 isbn = {1-59140-056-2},
 pages = {51--74},
 numpages = {24},
 url = {http://portal.acm.org/citation.cfm?id=887006.887010},
 acmid = {887010},
 publisher = {IGI Publishing},
 address = {Hershey, PA, USA},
}









@InBook{Kong:2004:IEC:123456.887010,
  author = 	{Kong, Wei-Chang},
  editor =	{Theerasak Thanasankit},
  title =	{E-commerce and cultural values - (InBook-num-in-chap)},
  chapter =	{9},
  year =	{2004},
  address =	{Hershey, PA, USA},
  publisher =	{IGI Publishing},
  url =		{http://portal.acm.org/citation.cfm?id=887006.887010},
  pages =	{51--74},
  numpages =	{24},
  acmid =	{887010},
  isbn =	{1-59140-056-2},
  number = 	 "",
  type = 	 "",
  month = 	 "",
  note = 	 "",
}


@Inbook{Kong:2005:IEC:887006.887010,
  author = 	{Kong, Wei-Chang},
  editor =	{Theerasak Thanasankit},
  title =	{E-commerce and cultural values (Inbook-text-in-chap)},
  chapter =	{The implementation of electronic commerce in SMEs in Singapore},
  year =	{2005},
  address =	{Hershey, PA, USA},
  publisher =	{IGI Publishing},
  url =		{http://portal.acm.org/citation.cfm?id=887006.887010},
  type =        {Chapter:},
  pages =	{51--74},
  numpages =	{24},
  acmid =	{887010},
  isbn =	{1-59140-056-2},
  number = 	 "",
  month = 	 "",
  note = 	 "",
}


@Inbook{Kong:2006:IEC:887006.887010,
  author = 	{Kong, Wei-Chang},
  editor =	{Theerasak Thanasankit},
  title =	{E-commerce and cultural values (Inbook-num chap)},
  chapter =	{22},
  year =	{2006},
  address =	{Hershey, PA, USA},
  publisher =	{IGI Publishing},
  url =		{http://portal.acm.org/citation.cfm?id=887006.887010},
  type =        {Chapter (in type field)},
  pages =	{51--74},
  numpages =	{24},
  acmid =	{887010},
  isbn =	{1-59140-056-2},
  number = 	 "",
  month = 	 "",
  note = 	 "",
}



@article{SaeediMEJ10,
            author = {Mehdi Saeedi and Morteza Saheb Zamani and Mehdi Sedighi},
            title = {A library-based synthesis methodology for reversible logic},
            journal = {Microelectron. J.},
            volume = {41},
            number = {4},
            month = apr,
            year = {2010},
            pages = {185--194},
}

@ARTICLE{SaeediJETC10,
            author = {Mehdi Saeedi and Morteza Saheb Zamani and Mehdi Sedighi and Zahra Sasanian},
            title = {Synthesis of Reversible Circuit Using Cycle-Based Approach},
            journal = {J. Emerg. Technol. Comput. Syst.},
            volume = {6},
            number = {4},
            month = dec,
            year = {2010}
            }

@article{Kirschmer:2010:AEI:1958016.1958018,
 author = {Kirschmer, Markus and Voight, John},
 title = {Algorithmic Enumeration of Ideal Classes for Quaternion Orders},
 journal = {SIAM J. Comput.},
 issue_date = {January 2010},
 volume = {39},
 number = {5},
 month = jan,
 year = {2010},
 issn = {0097-5397},
 pages = {1714--1747},
 numpages = {34},
 url = {http://dx.doi.org/10.1137/080734467},
 doi = {https://doi.org/10.1137/080734467},
 acmid = {1958018},
 publisher = {Society for Industrial and Applied Mathematics},
 address = {Philadelphia, PA, USA},
 keywords = {ideal classes, maximal orders, number theory, quaternion algebras},
}


@incollection{Hoare:1972:CIN:1243380.1243382,
 author = {Hoare, C. A. R.},
 title = {Chapter II: Notes on data structuring},
 booktitle = {Structured programming (incoll)},
 editor = {Dahl, O. J. and Dijkstra, E. W. and Hoare, C. A. R.},
 year = {1972},
 isbn = {0-12-200550-3},
 pages = {83--174},
 numpages = {92},
 url = {http://portal.acm.org/citation.cfm?id=1243380.1243382},
 acmid = {1243382},
 publisher = {Academic Press Ltd.},
 address = {London, UK, UK},
} 

@incollection{Lee:1978:TQA:800025.1198348,
 author = {Lee, Jan},
 title = {Transcript of question and answer session},
 booktitle = {History of programming languages I (incoll)},
 editor = {Wexelblat, Richard L.},
 year = {1981},
 isbn = {0-12-745040-8},
 pages = {68--71},
 numpages = {4},
 url = {http://doi.acm.org/10.1145/800025.1198348},
 doi = {http://doi.acm.org/10.1145/800025.1198348},
 acmid = {1198348},
 publisher = {ACM},
 address = {New York, NY, USA},
}

@incollection{Dijkstra:1979:GSC:1241515.1241518,
 author = {Dijkstra, E.},
 title = {Go to statement considered harmful},
 booktitle = {Classics in software engineering (incoll)},
 year = {1979},
 isbn = {0-917072-14-6},
 pages = {27--33},
 numpages = {7},
 url = {http://portal.acm.org/citation.cfm?id=1241515.1241518},
 acmid = {1241518},
 publisher = {Yourdon Press},
 address = {Upper Saddle River, NJ, USA},
} 

@incollection{Wenzel:1992:TVA:146022.146089,
 author = {Wenzel, Elizabeth M.},
 title = {Three-dimensional virtual acoustic displays},
 booktitle = {Multimedia interface design (incoll)},
 year = {1992},
 isbn = {0-201-54981-6},
 pages = {257--288},
 numpages = {32},
 url = {http://portal.acm.org/citation.cfm?id=146022.146089},
 doi = {10.1145/146022.146089},
 acmid = {146089},
 publisher = {ACM},
 address = {New York, NY, USA},
}

@incollection{Mumford:1987:MES:54905.54911,
 author = {Mumford, E.},
 title = {Managerial expert systems and organizational change: some critical research issues},
 booktitle = {Critical issues in information systems research (incoll)},
 year = {1987},
 isbn = {0-471-91281-6},
 pages = {135--155},
 numpages = {21},
 url = {http://portal.acm.org/citation.cfm?id=54905.54911},
 acmid = {54911},
 publisher = {John Wiley \& Sons, Inc.},
 address = {New York, NY, USA},
} 

@book{McCracken:1990:SSC:575315,
 author = {McCracken, Daniel D. and Golden, Donald G.},
 title = {Simplified Structured COBOL with Microsoft/MicroFocus COBOL},
 year = {1990},
 isbn = {0471514071},
 publisher = {John Wiley \& Sons, Inc.},
 address = {New York, NY, USA},
} 


@book {MR781537,
    AUTHOR = {H{\"o}rmander, Lars},
     TITLE = {The analysis of linear partial differential operators. {III}},
    SERIES = {Grundlehren der Mathematischen Wissenschaften [Fundamental
              Principles of Mathematical Sciences]},
    VOLUME = {275},
      NOTE = {Pseudodifferential operators},  
PUBLISHER = {Springer-Verlag},
   ADDRESS = {Berlin, Germany},
      YEAR = {1985},
     PAGES = {viii+525},
      ISBN = {3-540-13828-5},
   MRCLASS = {35-02 (35Sxx 47G05 58G15)},
  MRNUMBER = {781536 (87d:35002a)},
MRREVIEWER = {Min You Qi},
}

@book {MR781536,
    AUTHOR = {H{\"o}rmander, Lars},
     TITLE = {The analysis of linear partial differential operators. {IV}},
    SERIES = {Grundlehren der Mathematischen Wissenschaften [Fundamental
              Principles of Mathematical Sciences]},
    VOLUME = {275},
      NOTE = {Fourier integral operators},  
PUBLISHER = {Springer-Verlag},
   ADDRESS = {Berlin, Germany},
      YEAR = {1985},
     PAGES = {vii+352},
      ISBN = {3-540-13829-3},
   MRCLASS = {35-02 (35Sxx 47G05 58G15)},
  MRNUMBER = {781537 (87d:35002b)},
MRREVIEWER = {Min You Qi},
}


@InProceedings{Adya-01,
  author        = {A. Adya and P. Bahl and J. Padhye and A.Wolman and L. Zhou},
  title         = {A multi-radio unification protocol for {IEEE} 802.11 wireless networks},
  booktitle     = {Proceedings of the IEEE 1st International Conference on Broadnets Networks (BroadNets'04)},
  publisher     = "IEEE",
  address       = "Los Alamitos, CA",		  
  year          = {2004},
  pages         = "210--217"
}

@article{Akyildiz-01,
  author        = {I. F. Akyildiz and W. Su and Y. Sankarasubramaniam and E. Cayirci},
  title         = {Wireless Sensor Networks: A Survey},
  journal       = {Comm. ACM},
  volume        = 38,
  number        = "4",
  year          = {2002},
  pages         = "393--422"
}

@article{Akyildiz-02,
  author        = {I. F. Akyildiz and T. Melodia and K. R. Chowdhury},
  title         = {A Survey on Wireless Multimedia Sensor Networks},
  journal       = {Computer Netw.},
  volume        = 51,
  number        = "4",
  year          = {2007},
  pages         = "921--960"
}

@InProceedings{Bahl-02,
  author        = {P. Bahl and R. Chancre and J. Dungeon},
  title         = {{SSCH}: Slotted Seeded Channel Hopping for Capacity Improvement in {IEEE} 802.11 Ad-Hoc Wireless Networks},
  booktitle     = {Proceeding of the 10th International Conference on Mobile Computing and Networking (MobiCom'04)},
  publisher     = "ACM",
  address       = "New York, NY",		  
  year          = {2004},
  pages         = "112--117"
}

@misc{CROSSBOW,
  key       = {CROSSBOW},
  title     = {{XBOW} Sensor Motes Specifications},
  note      = {http://www.xbow.com},
  year      = 2008
}

@article{Culler-01,
  author        = {D. Culler and D. Estrin and M. Srivastava},
  title         = {Overview of Sensor Networks},
  journal       = {IEEE Comput.},
  volume        = 37,
  number        = "8 (Special Issue on Sensor Networks)",
  publisher     = "IEEE",
  address       = "Los Alamitos, CA",		  
  year          = {2004},
  pages         = "41--49"
}

@misc{Harvard-01,
    key         = {Harvard CodeBlue},
    title       = {{CodeBlue}: Sensor Networks for Medical Care},
    note        = {http://www.eecs.harvard.edu/mdw/ proj/codeblue/},
    year        = 2008
}

@InProceedings{Natarajan-01,
    author      = {A. Natarajan and M. Motani and B. de Silva and K. Yap and K. C. Chua},
    title       = {Investigating Network Architectures for Body Sensor Networks},
    booktitle   = {Network Architectures},
    editor      = {G. Whitcomb and P. Neece},
    publisher   = "Keleuven Press",
    address     = "Dayton, OH",		  
    year        = {2007},
    pages       = "322--328",
    eprint      = "960935712",
    primaryclass = "cs",
}

@techreport{Tzamaloukas-01,
  author        = {A. Tzamaloukas and J. J. Garcia-Luna-Aceves},
  title         = {Channel-Hopping Multiple Access},
  number =        {I-CA2301},
  institution =   {Department of Computer Science, University of California},
  address =       {Berkeley, CA},
  year          = {2000}
}

@BOOK{Zhou-06,
  author        = {G. Zhou and J. Lu and C.-Y. Wan and M. D. Yarvis and J. A. Stankovic},
  title         = {Body Sensor Networks},
  publisher     = "MIT Press",
  address       = "Cambridge, MA",		  
  year          = {2008}
}

@mastersthesis{ko94,
author = "Jacob Kornerup",
title = "Mapping Powerlists onto Hypercubes",
school = "The University of Texas at Austin",
note = "(In preparation)",
year = "1994"}

@PhdThesis{gerndt:89,
  author =       "Michael Gerndt",
  title =        "Automatic Parallelization for Distributed-Memory
                  Multiprocessing Systems",
  school =       "University of Bonn",
  year =         1989,
  address =      "Bonn, Germany",
  month =        dec
}

@article{6:1:1,
author = "J. E. {Archer, Jr.} and R. Conway and F. B. Schneider",
title = "User recovery and reversal in interactive systems",
journal = "ACM Trans. Program. Lang. Syst.",
volume =  "6",
number = "1",
month = jan,
year = 1984,
pages = "1--19"}

@article{7:1:137,
author = "D. D. Dunlop and V. R. Basili",
title = "Generalizing specifications for uniformly implemented loops",
journal = "ACM Trans. Program. Lang. Syst.",
volume =  "7",
number = "1",
month = jan,
year = 1985,
pages = "137--158"}

@article{7:2:183,
author = "J. Heering and P. Klint",
title = "Towards monolingual programming environments",
journal = "ACM Trans. Program. Lang. Syst.",
volume =  "7",
number = "2",
month = apr,
year = 1985,
pages = "183--213"}

@book{knuth:texbook,
author = "Donald E. Knuth",
title = "The {\TeX{}book}",
publisher = "Addison-Wesley",
address = "Reading, MA.",
year = 1984}

@article{6:3:380,
author = "E. Korach and D.  Rotem and N. Santoro",
title = "Distributed algorithms for finding centers and medians in networks",
journal = "ACM Trans. Program. Lang. Syst.",
volume =  "6",
number = "3",
month = jul,
year = 1984,
pages = "380--401"}

@book{lamport:latex,
author = "Leslie Lamport",
title = "\it {\LaTeX}: A Document Preparation System",
publisher = "Addison-Wesley",
address = "Reading, MA.",
year = 1986}

@article{7:3:359,
author = "F. Nielson",
title = "Program transformations in a denotational setting",
journal = "ACM Trans. Program. Lang. Syst.",
volume =  "7",
number = "3",
month = jul,
year = 1985,
pages = "359--379"}

@misc{ps073006_2018_1157183,
  author       = {ps073006 and
                  Upulee Kanewala},
  title        = {MSU-STLab/METtester 1.0.0},
  month        = jan,
  year         = 2018,
  doi          = {10.5281/zenodo.1157183},
  url          = {https://doi.org/10.5281/zenodo.1157183}
}















\begin{thebibliography}{20}


\ifx \showCODEN    \undefined \def \showCODEN     #1{\unskip}     \fi
\ifx \showDOI      \undefined \def \showDOI       #1{#1}\fi
\ifx \showISBNx    \undefined \def \showISBNx     #1{\unskip}     \fi
\ifx \showISBNxiii \undefined \def \showISBNxiii  #1{\unskip}     \fi
\ifx \showISSN     \undefined \def \showISSN      #1{\unskip}     \fi
\ifx \showLCCN     \undefined \def \showLCCN      #1{\unskip}     \fi
\ifx \shownote     \undefined \def \shownote      #1{#1}          \fi
\ifx \showarticletitle \undefined \def \showarticletitle #1{#1}   \fi
\ifx \showURL      \undefined \def \showURL       {\relax}        \fi
\providecommand\bibfield[2]{#2}
\providecommand\bibinfo[2]{#2}
\providecommand\natexlab[1]{#1}
\providecommand\showeprint[2][]{arXiv:#2}

\bibitem[\protect\citeauthoryear{Arcuri}{Arcuri}{2010}]%
        {5477082}
\bibfield{author}{\bibinfo{person}{A. Arcuri}.}
  \bibinfo{year}{2010}\natexlab{}.
\newblock \showarticletitle{It Does Matter How You Normalise the Branch
  Distance in Search Based Software Testing}. In \bibinfo{booktitle}{{\em 2010
  Third International Conference on Software Testing, Verification and
  Validation}}. \bibinfo{pages}{205--214}.
\newblock
\showISSN{2159-4848}
\showDOI{%
\url{https://doi.org/10.1109/ICST.2010.17}}


\bibitem[\protect\citeauthoryear{Barr, Harman, McMinn, Shahbaz, and Yoo}{Barr
  et~al\mbox{.}}{2015}]%
        {6963470}
\bibfield{author}{\bibinfo{person}{E.~T. Barr}, \bibinfo{person}{M. Harman},
  \bibinfo{person}{P. McMinn}, \bibinfo{person}{M. Shahbaz}, {and}
  \bibinfo{person}{S. Yoo}.} \bibinfo{year}{2015}\natexlab{}.
\newblock \showarticletitle{The Oracle Problem in Software Testing: A Survey}.
\newblock \bibinfo{journal}{{\em IEEE Transactions on Software Engineering\/}}
  \bibinfo{volume}{41}, \bibinfo{number}{5} (\bibinfo{date}{May}
  \bibinfo{year}{2015}), \bibinfo{pages}{507--525}.
\newblock
\showISSN{0098-5589}
\showDOI{%
\url{https://doi.org/10.1109/TSE.2014.2372785}}


\bibitem[\protect\citeauthoryear{Batra and Sengupta}{Batra and
  Sengupta}{2011}]%
        {10.1007/978-3-642-19423-8_19}
\bibfield{author}{\bibinfo{person}{Gagandeep Batra} {and}
  \bibinfo{person}{Jyotsna Sengupta}.} \bibinfo{year}{2011}\natexlab{}.
\newblock \showarticletitle{An Efficient Metamorphic Testing Technique Using
  Genetic Algorithm}. In \bibinfo{booktitle}{{\em Information Intelligence,
  Systems, Technology and Management}},
  \bibfield{editor}{\bibinfo{person}{Sumeet Dua}, \bibinfo{person}{Sartaj
  Sahni}, {and} \bibinfo{person}{D.~P. Goyal}} (Eds.).
  \bibinfo{publisher}{Springer Berlin Heidelberg}, \bibinfo{address}{Berlin,
  Heidelberg}, \bibinfo{pages}{180--188}.
\newblock
\showISBNx{978-3-642-19423-8}


\bibitem[\protect\citeauthoryear{Chen, Cai, Liu, Liu, Wei, and Liu}{Chen
  et~al\mbox{.}}{2012a}]%
        {6528673}
\bibfield{author}{\bibinfo{person}{Leilei Chen}, \bibinfo{person}{Lizhi Cai},
  \bibinfo{person}{Jiang Liu}, \bibinfo{person}{Zhenyu Liu},
  \bibinfo{person}{Shiyan Wei}, {and} \bibinfo{person}{Pan Liu}.}
  \bibinfo{year}{2012}\natexlab{a}.
\newblock \showarticletitle{An optimized method for generating cases of
  metamorphic testing}. In \bibinfo{booktitle}{{\em 2012 6th International
  Conference on New Trends in Information Science, Service Science and Data
  Mining (ISSDM2012)}}. \bibinfo{pages}{439--443}.
\newblock


\bibitem[\protect\citeauthoryear{Chen}{Chen}{2015}]%
        {Chen:2015:MTS:2819261.2819278}
\bibfield{author}{\bibinfo{person}{Tsong~Yueh Chen}.}
  \bibinfo{year}{2015}\natexlab{}.
\newblock \showarticletitle{Metamorphic Testing: A Simple Method for
  Alleviating the Test Oracle Problem}. In \bibinfo{booktitle}{{\em Proceedings
  of the 10th International Workshop on Automation of Software Test}} {\em
  (\bibinfo{series}{AST '15})}. \bibinfo{publisher}{IEEE Press},
  \bibinfo{address}{Piscataway, NJ, USA}, \bibinfo{pages}{53--54}.
\newblock
\showURL{%
\url{http://dl.acm.org/citation.cfm?id=2819261.2819278}}


\bibitem[\protect\citeauthoryear{Chen, Kuo, Liu, and Tang}{Chen
  et~al\mbox{.}}{2004}]%
        {DBLP:ChenKLT04}
\bibfield{author}{\bibinfo{person}{Tsong~Yueh Chen},
  \bibinfo{person}{Fei{-}Ching Kuo}, \bibinfo{person}{Ying Liu}, {and}
  \bibinfo{person}{Antony Tang}.} \bibinfo{year}{2004}\natexlab{}.
\newblock \showarticletitle{Metamorphic Testing and Testing with Special
  Values}. In \bibinfo{booktitle}{{\em 4th {IEEE} International Workshop on
  Source Code Analysis and Manipulation {(SCAM} 2004), 15-16 September 2004,
  Chicago, IL, {USA}}}. \bibinfo{pages}{128--134}.
\newblock


\bibitem[\protect\citeauthoryear{Chen, Kuo, Towey, and Zhou}{Chen
  et~al\mbox{.}}{2012b}]%
        {6319263}
\bibfield{author}{\bibinfo{person}{T.~Y. Chen}, \bibinfo{person}{F.~C. Kuo},
  \bibinfo{person}{D. Towey}, {and} \bibinfo{person}{Z.~Q. Zhou}.}
  \bibinfo{year}{2012}\natexlab{b}.
\newblock \showarticletitle{Metamorphic Testing: Applications and Integration
  with Other Methods: Tutorial Synopsis}. In \bibinfo{booktitle}{{\em 2012 12th
  International Conference on Quality Software}}. \bibinfo{pages}{285--288}.
\newblock
\showISSN{1550-6002}
\showDOI{%
\url{https://doi.org/10.1109/QSIC.2012.21}}


\bibitem[\protect\citeauthoryear{Dong, Guo, and Zhang}{Dong
  et~al\mbox{.}}{2013}]%
        {6615286}
\bibfield{author}{\bibinfo{person}{Guowei Dong}, \bibinfo{person}{Tao Guo},
  {and} \bibinfo{person}{Puhan Zhang}.} \bibinfo{year}{2013}\natexlab{}.
\newblock \showarticletitle{Security assurance with program path analysis and
  metamorphic testing}. In \bibinfo{booktitle}{{\em 2013 IEEE 4th International
  Conference on Software Engineering and Service Science}}.
  \bibinfo{pages}{193--197}.
\newblock
\showISSN{2327-0586}
\showDOI{%
\url{https://doi.org/10.1109/ICSESS.2013.6615286}}


\bibitem[\protect\citeauthoryear{Fraser and Arcuri}{Fraser and Arcuri}{2011}]%
        {Fraser:2011:EAT:2025113.2025179}
\bibfield{author}{\bibinfo{person}{Gordon Fraser} {and} \bibinfo{person}{Andrea
  Arcuri}.} \bibinfo{year}{2011}\natexlab{}.
\newblock \showarticletitle{EvoSuite: Automatic Test Suite Generation for
  Object-oriented Software}. In \bibinfo{booktitle}{{\em Proceedings of the
  19th ACM SIGSOFT Symposium and the 13th European Conference on Foundations of
  Software Engineering}} {\em (\bibinfo{series}{ESEC/FSE '11})}.
  \bibinfo{publisher}{ACM}, \bibinfo{address}{New York, NY, USA},
  \bibinfo{pages}{416--419}.
\newblock
\showISBNx{978-1-4503-0443-6}
\showDOI{%
\url{https://doi.org/10.1145/2025113.2025179}}


\bibitem[\protect\citeauthoryear{Gotlieb and Botella}{Gotlieb and
  Botella}{2003}]%
        {1245319}
\bibfield{author}{\bibinfo{person}{A. Gotlieb} {and} \bibinfo{person}{B.
  Botella}.} \bibinfo{year}{2003}\natexlab{}.
\newblock \showarticletitle{Automated metamorphic testing}. In
  \bibinfo{booktitle}{{\em Proceedings 27th Annual International Computer
  Software and Applications Conference. COMPAC 2003}}. \bibinfo{pages}{34--40}.
\newblock
\showISSN{0730-3157}
\showDOI{%
\url{https://doi.org/10.1109/CMPSAC.2003.1245319}}


\bibitem[\protect\citeauthoryear{Korel}{Korel}{1990}]%
        {Korel:1990:AST:101747.101755}
\bibfield{author}{\bibinfo{person}{B. Korel}.} \bibinfo{year}{1990}\natexlab{}.
\newblock \showarticletitle{Automated Software Test Data Generation}.
\newblock \bibinfo{journal}{{\em IEEE Trans. Softw. Eng.\/}}
  \bibinfo{volume}{16}, \bibinfo{number}{8} (\bibinfo{date}{Aug.}
  \bibinfo{year}{1990}), \bibinfo{pages}{870--879}.
\newblock
\showISSN{0098-5589}
\showDOI{%
\url{https://doi.org/10.1109/32.57624}}


\bibitem[\protect\citeauthoryear{Ma, Offutt, and Kwon}{Ma
  et~al\mbox{.}}{2005}]%
        {Ma:2005:MAC:1077303.1077304}
\bibfield{author}{\bibinfo{person}{Yu-Seung Ma}, \bibinfo{person}{Jeff Offutt},
  {and} \bibinfo{person}{Yong~Rae Kwon}.} \bibinfo{year}{2005}\natexlab{}.
\newblock \showarticletitle{MuJava: An Automated Class Mutation System:
  Research Articles}.
\newblock \bibinfo{journal}{{\em Softw. Test. Verif. Reliab.\/}}
  \bibinfo{volume}{15}, \bibinfo{number}{2} (\bibinfo{date}{June}
  \bibinfo{year}{2005}), \bibinfo{pages}{97--133}.
\newblock
\showISSN{0960-0833}
\showDOI{%
\url{https://doi.org/10.1002/stvr.v15:2}}


\bibitem[\protect\citeauthoryear{McMinn}{McMinn}{2004}]%
        {McMinn:2004:SST:1077276.1077279}
\bibfield{author}{\bibinfo{person}{Phil McMinn}.}
  \bibinfo{year}{2004}\natexlab{}.
\newblock \showarticletitle{Search-based Software Test Data Generation: A
  Survey: Research Articles}.
\newblock \bibinfo{journal}{{\em Softw. Test. Verif. Reliab.\/}}
  \bibinfo{volume}{14}, \bibinfo{number}{2} (\bibinfo{date}{June}
  \bibinfo{year}{2004}), \bibinfo{pages}{105--156}.
\newblock
\showISSN{0960-0833}
\showDOI{%
\url{https://doi.org/10.1002/stvr.v14:2}}


\bibitem[\protect\citeauthoryear{Murphy, Kaiser, Hu, and Wu}{Murphy
  et~al\mbox{.}}{2008}]%
        {unknown}
\bibfield{author}{\bibinfo{person}{Christian Murphy}, \bibinfo{person}{Gail
  Kaiser}, \bibinfo{person}{Lifeng Hu}, {and} \bibinfo{person}{Leon Wu}.}
  \bibinfo{year}{2008}\natexlab{}.
\newblock \showarticletitle{Properties of Machine Learning Applications for Use
  in Metamorphic Testing.} \bibinfo{pages}{867--872}.
\newblock


\bibitem[\protect\citeauthoryear{ps073006 and Kanewala}{ps073006 and
  Kanewala}{2018}]%
        {ps073006_2018_1157183}
\bibfield{author}{\bibinfo{person}{ps073006} {and} \bibinfo{person}{Upulee
  Kanewala}.} \bibinfo{year}{2018}\natexlab{}.
\newblock \bibinfo{title}{MSU-STLab/METtester 1.0.0}.
\newblock   (\bibinfo{date}{Jan.} \bibinfo{year}{2018}).
\newblock
\showDOI{%
\url{https://doi.org/10.5281/zenodo.1157183}}


\bibitem[\protect\citeauthoryear{Rojas, Campos, Vivanti, Fraser, and
  Arcuri}{Rojas et~al\mbox{.}}{2015}]%
        {Rojas2015}
\bibfield{author}{\bibinfo{person}{Jos{\'e}~Miguel Rojas},
  \bibinfo{person}{Jos{\'e} Campos}, \bibinfo{person}{Mattia Vivanti},
  \bibinfo{person}{Gordon Fraser}, {and} \bibinfo{person}{Andrea Arcuri}.}
  \bibinfo{year}{2015}\natexlab{}.
\newblock \bibinfo{booktitle}{{\em Combining Multiple Coverage Criteria in
  Search-Based Unit Test Generation}}.
\newblock \bibinfo{publisher}{Springer International Publishing},
  \bibinfo{address}{Cham}, \bibinfo{pages}{93--108}.
\newblock
\showISBNx{978-3-319-22183-0}
\showDOI{%
\url{https://doi.org/10.1007/978-3-319-22183-0_7}}


\bibitem[\protect\citeauthoryear{Segura, Hierons, Benavides, and
  Ruiz-Cortés}{Segura et~al\mbox{.}}{2011}]%
        {SEGURA2011245}
\bibfield{author}{\bibinfo{person}{Sergio Segura}, \bibinfo{person}{Robert~M.
  Hierons}, \bibinfo{person}{David Benavides}, {and} \bibinfo{person}{Antonio
  Ruiz-Cortés}.} \bibinfo{year}{2011}\natexlab{}.
\newblock \showarticletitle{Automated metamorphic testing on the analyses of
  feature models}.
\newblock \bibinfo{journal}{{\em Information and Software Technology\/}}
  \bibinfo{volume}{53}, \bibinfo{number}{3} (\bibinfo{year}{2011}),
  \bibinfo{pages}{245 -- 258}.
\newblock
\showISSN{0950-5849}
\showDOI{%
\url{https://doi.org/10.1016/j.infsof.2010.11.002}}


\bibitem[\protect\citeauthoryear{Sen and Agha}{Sen and Agha}{2006}]%
        {conf/cav/SenA06}
\bibfield{author}{\bibinfo{person}{Koushik Sen} {and} \bibinfo{person}{Gul
  Agha}.} \bibinfo{year}{2006}\natexlab{}.
\newblock \showarticletitle{CUTE and jCUTE: Concolic Unit Testing and Explicit
  Path Model-Checking Tools}. In \bibinfo{booktitle}{{\em CAV}},
  \bibfield{editor}{\bibinfo{person}{Thomas Ball} {and}
  \bibinfo{person}{Robert~B. Jones}} (Eds.). \bibinfo{pages}{419--423}.
\newblock


\bibitem[\protect\citeauthoryear{Weyuker}{Weyuker}{1982}]%
        {Weyukerarticle}
\bibfield{author}{\bibinfo{person}{Elaine Weyuker}.}
  \bibinfo{year}{1982}\natexlab{}.
\newblock \showarticletitle{On Testing Non-Testable Programs}.
\newblock   \bibinfo{volume}{25} (\bibinfo{date}{11} \bibinfo{year}{1982}).
\newblock


\bibitem[\protect\citeauthoryear{Wu, Xiao-Chun, Jiang-Jun, and Hui-Min}{Wu
  et~al\mbox{.}}{2005}]%
        {article}
\bibfield{author}{\bibinfo{person}{Peng Wu}, \bibinfo{person}{SHI Xiao-Chun},
  \bibinfo{person}{TANG Jiang-Jun}, {and} \bibinfo{person}{LIN Hui-Min}.}
  \bibinfo{year}{2005}\natexlab{}.
\newblock \showarticletitle{Metamorphic Testing and Special Case Testing: A
  Case Study}.
\newblock   \bibinfo{volume}{16} (\bibinfo{date}{07} \bibinfo{year}{2005}).
\newblock


\end{thebibliography}

\end{document}